\documentstyle[aps,eqsecnum]{revtex}
\begin{document}
\draft

\title{The noise in gravitational-wave detectors and other classical-force
measurements is not influenced by test-mass quantization
}

\author{Vladimir B.\ Braginsky$^1$, Mikhail L.\ Gorodetsky$^1$,
Farid Ya.\ Khalili$^1$, Andrey B.\ Matsko$^2$, Kip S.\ Thorne$^3$, and
Sergey P.\ Vyatchanin$^1$}

\address{$^1$Physics Faculty, Moscow State University, Moscow Russia}
\address{$^2$Department of Physics, Texas A\&M University, College Station, TX
77843-4242}
\address{$^3$Theoretical Astrophysics, California Institute of
Technology, Pasadena, CA 91125}

\date{Received 2 September 2001; Revised 4 July 2002 and 1 December 2002}

\maketitle

\begin{abstract}

It is shown that photon shot noise and radiation-pressure back-action noise
are the sole forms of quantum noise in interferometric gravitational wave
detectors that operate near or below the standard quantum limit,
if one filters the interferometer output appropriately.
No
additional noise arises from the 
test masses' initial quantum state or from
reduction of the test-mass state due to measurement of the
interferometer output or from the uncertainty principle associated with
the test-mass state.
Two features of interferometers are central to these
conclusions: (i) The interferometer output (the photon number flux $\hat
{\cal N}(t)$ entering the final photodetector) commutes with itself at
different times in the Heisenberg Picture, $[\hat {\cal N}(t),\hat {\cal
N}(t')] = 0$ and thus can be regarded as classical. (ii) This number flux is
linear 
to high accuracy
in the test-mass initial position and momentum operators $\hat x_o$
and $\hat p_o$, 
and those operators influence the measured photon flux $\hat {\cal N}(t)$
in manners that
can easily be removed by filtering.  For example, in most interferometers
$\hat x_o$ and $\hat p_o$ appear in $\hat{\cal N}(t)$ only 
at the test masses' $\sim 1$~Hz pendular swinging freqency and their
influence is 
removed when the output data are high-pass filtered to
get rid of noise below $\sim 10$ Hz.

The test-mass operators $\hat x_o$ and $\hat p_o$ contained in the unfiltered
output $\hat {\cal N}(t)$ make a nonzero contribution to the commutator
$[\hat {\cal N}(t), \hat {\cal N}(t')]$.  That contribution is precisely
cancelled by a nonzero commutation of the photon shot noise and
radiation-pressure noise, which also are contained in $\hat {\cal N}(t)$.
This cancellation of commutators is responsible for the fact that it is
possible to derive an interferometer's standard quantum limit from test-mass
considerations, and independently from photon-noise considerations, and get
identically the same result.

These conclusions are all true for a far wider class of measurements than just
gravitational-wave interferometers.  To elucidate them, this paper
presents a series of idealized thought experiments that are free from
the complexities of real measuring systems.

\end{abstract}

\narrowtext
\twocolumn

\section{Questions to be analyzed and summary of answers}
\label{sec:Questions}

It has long been known that the Heisenberg uncertainty principle imposes a
``standard quantum limit'' (SQL) on high-precision measurements
\cite{sql,qnd1,QuantumMeasurement}. This SQL can be circumvented by using
``quantum nondemolition'' (QND) techniques
\cite{qnd1,QuantumMeasurement,qnd2,toys,VMZ,VM1,VM2,VL}.

For broad-band interferometric gravitational-wave detectors the SQL is a
limiting (single-sided) spectral density

\begin{equation}
  S_h(f) = {8\hbar\over m(2\pi f)^2 L^2}\; \label{sql}
\end{equation}
for the gravitational-wave field $h(t)$ \cite{caves1,300years}.  Here $\hbar$
is Planck's constant divided by $2\pi$, $m$ is the mass of each of the
interferometer's four test masses, $L$ is the interferometer's arm length, and $f$
is frequency.

This SQL firmly constrains the sensitivity of all conventional
interferometers (interferometers with the same optical topology as LIGO's
first-generation gravitational-wave detectors) 
\cite{pace,kimble}.  
LIGO's
second-generation interferometers (LIGO-II; ca.\ 2008) are expected to reach
this SQL for their $m= 40$ kg test masses in the vicinity of $f\sim 100$ Hz
\cite{lsc}, and may even beat it by a modest amount thanks to a ``signal
recycling mirror'' that converts them from conventional interferometers into
QND devices \cite{BC1,BC2,BC3}.  LIGO-III interferometers are likely to
beat the SQL by a factor $\sim 4$ or more; 
see, e.g., \cite{kimble}.  

In the 
research and development 
for LIGO-II
interferometers \cite{lsc,BC1,BC2,BC3} 
and in the attempts to invent strongly QND
LIGO-III interferometers
\cite{unruh,OpticalBar,SymphotonicState,DualResonator,LNRigidity,FDRigidity,SpeedMeter,kimble}, 
it is
important to understand clearly the physical nature of the quantum noise
which imposes the SQL, and to be able to compute with confidence the spectral
density of this quantum noise for various interferometer designs.  These
issues are the subject of this paper.

There are two standard ways to derive the gravitational-wave SQL (\ref{sql}),
and correspondingly two different viewpoints on it. The first derivation
\cite{caves1,hollenhorst} 
focuses on the quantum mechanics of the
interferometer's test masses and ignores the interferometer's other details.
In the simplest version of this derivation, one imagines a sequence of
instantaneous measurements of the difference

\begin{equation}
\hat x \equiv (\hat x_1 - \hat x_2) - (\hat x_3 - \hat x_4)
\label{xDef}
\end{equation}
of the center-of-mass positions of the four test masses, and from this
measurement sequence one infers the changes of $x$ and thence the time
varying gravitational-wave field $h(t) = x(t)/L$.  At time $t$ immediately
after one of the measurements, the test masses' reduced state has position
variance $[\Delta x(t)]^2$ no smaller than the measurement's accuracy.
During the time interval $\tau = t'-t$ between this measurement and the next,
the test masses are free, so $\hat x(t)$ evolves as the position of a free
particle with mass 
\begin{equation}
\mu = m/4 
\label{muDef}
\end{equation}
[the reduced mass of the four-body system
with relative position (\ref{xDef})].  
The Heisenberg-Picture commutation relations for a free particle

\begin{equation}
  [\hat x(t), \hat x({t'})] = {i\hbar (t'-t)\over \mu} =
  {4i\hbar \tau\over m}
\label{TMCommutator}
\end{equation}
imply that, whatever may be the state of the test masses, the variance
$[\Delta x({t'})]^2$ of $\hat x$ just before the next measurement must satisfy
the Heisenberg uncertainty relation

\begin{equation}
  \Delta x(t) \Delta x(t') \ge {\hbar |t-t'|\over 2\mu} =
    {2\hbar \tau\over m}\;.
\label{xUncertainty}
\end{equation}
The accuracy with which the change of $x$ between $t$ and $t'$ can be
measured is no better than the value obtained by setting $\Delta x(t) =
\Delta x(t')$, and in classical language that accuracy is related to the
minimum possible spectral density of the noise at frequency 
$f\simeq
1/\pi\tau$ by $\Delta x(t) = \Delta x(t') \simeq \sqrt{S_h(f) /\tau}$.
Simple algebra then gives expression (\ref{sql}) for the SQL
of $S_h(f)$.  A more sophisticated analysis \cite{caves1}, based on
measurements that are continuous rather than discrete and on a nonunitary
Feynman-path-integral evolution of the test-mass state
\cite{caves2,mensky2}, 
gives
precisely the SQL (\ref{sql}).

The second derivation of the SQL \cite{caves3,caves4} ignores the quantum
mechanics of the test mass, and focuses instead on that of the laser light
which monitors the test-mass motion.  The light produces two kinds of noise:
photon shot noise, which gets superposed on the output gravitational-wave
signal, and radiation-pressure fluctuations, which produce a random
back-action force on the test masses, thereby influencing their position
evolution and thence the interferometer output.  In an ideal, SQL-limited
interferometer, both noises --- shot and radiation-pressure --- arise from
quantum electrodynamic vacuum fluctuations that enter the interferometer
through its dark port and superpose on the highly classical laser light
\cite{caves3,caves4}.  The radiation-pressure spectral density is
proportional to the laser-light power $P$, the shot-noise spectral density is
proportional to $1/P$, and their product is independent of $P$ and is
constrained by the uncertainty principle for light (or equivalently by the
electromagnetic field commutation relations) to be no smaller than
\begin{equation}
S_x S_F = \hbar^2
\label{LightUncertainty}
\end{equation}
[cf.\ Eqs.\ (6.7) and (6.17) of \cite{QuantumMeasurement} in which there is
a factor 1/4 on the right side because Ref.\ \cite{QuantumMeasurement} uses
a double-sided spectral density, while the present paper uses the
gravity-wave community's single-sided convention].  In Eq.\
(\ref{LightUncertainty})
$S_x(f)$ is the spectral density of the shot noise that is superposed on the
interferometer's output position signal $x(t)$, 
$S_F(f)$ is the spectral
density of the radiation-pressure force that acts on the test-mass
center-of-mass degree of freedom $x$, and we have assumed that the shot
noise and radiation-pressure force are uncorrelated as is the case for
conventional (LIGO-I type) interferometers \cite{kimble,BC1,BC2,BC3}.
At frequency $f$ the test mass
responds to the Fourier component $\tilde F(f)$ of the force with a position
change 
$\tilde x(f) = - \tilde F(f)/[\mu(2\pi f)^2]$, and correspondingly
the net gravitational-wave noise is
\begin{equation}
  S_h(f) = {1\over L^2}\left(S_x + {S_F\over \mu^2(2\pi f)^4}\right)\;.
  \label{NetLightNoise}
\end{equation}
By combining Eqs.\ (\ref{LightUncertainty}), (\ref{NetLightNoise}) 
and (\ref{muDef}), 
we obtain the SQL (\ref{sql}) for a conventional interferometer, e.g.\ LIGO-I.

In view of these two very different derivations of the SQL, test-mass
quantization and light quantization, three questions arise: (i) Are the
test-mass quantization and the light quantization just two different
viewpoints on the same physics?--- in which case the correct SQL is Eq.\
(\ref{sql}). Or are they fully or partially independent effects? --- in which
case we would expect their noises to add, causing the true SQL for 
$S_h$ to be larger
by, perhaps, a factor 2 (and thence the event rate in an SQL-limited
interferometer
to be reduced by a factor $\sim(\sqrt2)^3 \simeq 3$). (ii) How should one
compute the quantum noise in candidate designs for the QND LIGO-II
and LIGO-III interferometers? One inevitably must pay close
attention to the behavior of the light (and thus also its quantization),
since the optical configuration will differ markedly from one candidate
design to another.  Must one also pay close attention to the quantum
mechanics of the test masses, including their commutation relation
(\ref{TMCommutator}) and the continual reduction of their state as
information about them is continually put onto the light's modulations and
then measured?  (iii) Similarly, how should one design a QND interferometer?
Need one adjust one's design so as to drive both the light's noise and the
test-mass noise below the SQL?

As we shall show, the answers are these: (ii) The test-mass quantization is
irrelevant to the interferometer's noise
and correspondingly test-mass state reduction is irrelevant, 
if one
filters the output data appropriately. (For 
interferometers with conventional optical topology such as LIGO-I,
it is sufficient to discard all data near the test masses' $\sim 1$ Hz
swinging frequency.) 
Therefore, 
one can ignore test-mass quantization and state reduction when computing
the noise of a candidate interferometer.
(iii) Similarly, one can ignore
the test mass's quantum noise
when designing a QND interferometer that beats the SQL. One need only
pay attention to the light's quantum noise, and in principle, by manipulating
the light appropriately 
(and filtering the output data appropriately), 
one can circumvent the SQL completely.
(i) Correspondingly, the SQL
(\ref{sql}) as derived from light quantization is precisely correct; there is
no extra factor 2 caused by test-mass quantization. 
[The fact that one can
also derive the SQL from test-mass quantization is a result of an intimate
connection between the uncertainty principles for a measured system (the
test masses in our case) and the system that makes the measurement (the light).
We shall elucidate this intimate connection from one viewpoint at the end
of Sec.\ \ref{ClassicalSQL}.  
From another viewpoint, it is due to the fact that the
commutator $[\hat x(t), \hat x(t')]$, which underlies the test-mass
derivation (\ref{TMCommutator}), (\ref{xUncertainty}) of the SQL, 
also underlies the derivation of the measuring light's 
uncertainty relation 
(\ref{LightUncertainty}); see the role of the generalized susceptibility
$\chi(t,t') = (1/i\hbar)[\hat x(t'),\hat x(t)]$ in Sec.\ 6.3 of Ref.\
\cite{QuantumMeasurement}.] 

Central to our answers (i), (ii) and (iii) is the fact that an
interferometric gravitational-wave detector does {\it not} monitor the
time-evolving test-mass position $\hat x(t)$.  Rather, it only monitors {\it
classical changes} in $\hat x(t)$ induced by the classical gravitational-wave
field $h(t)$ and other classical\footnote{All these forces --- 
gravitational-wave, thermal, seismic, etc.\ --- actually do have a quantum
component, but in practice their levels of excitation are so large that 
we can regard them as classical.}
forces (thermal, seismic, ...) acting on the test
masses, and it does so without extracting information about the actual
quantized position $\hat x(t)$.  The detector has a classical input [$h(t)$]
and a classical output [$h(t)$ contaminated by noise that 
(as we shall see) commutes with
itself at different times and that therefore can be regarded as a
time-evolving c-number].  The quantum properties of the test masses and the
light are merely intermediaries through which the classical signal must pass.
This would not be the case for a device designed to make a sequence of
absolute measurements of the quantum mechanical position $\hat x(t)$.

Our answers (i), (ii), (iii) hold true for a far wider range of measuring devices than just
interferometric gravitational-wave detectors. They hold quite generally for
any well-designed device that measures a classical
force acting on any quantum mechanical system.
In particular, they remain true if the device makes measurements that are {\it linear}
in the sense of Appendix \ref{app:LinearMeasurements}, and one filters
the device's output to remove all information at the natural frequencies of the
quantum system's dynamics (e.g.\ at its eigenfrequency if the quantum system
is a harmonic oscillator).

In Sec.\ \ref{sec:Pedagogy} we will elucidate these answers by considering
pedagogical examples of idealized devices that make discrete, quick
measurements on a test mass.  These examples will reveal two central
underpinnings of our answers: (a) the vanishing of the measurement's ``output
commutators'' --- i.e., the commutators of the observables (Hermitian
operators) that represent the entries in the output data stream, and (b) a
data-processing procedure that removes from the data all influence of the
test-mass quantum observables (initial position $\hat x_o$ and initial
momentum $\hat p_o$).  Our examples will also elucidate two strategies for
beating the SQL: (A) put the measuring apparatus (``meters'') into specially
chosen initial states (the analog of squeezed states), and (B) measure a
wisely chosen linear combination of position and momentum for the test mass
and thereby remove the effects of the meters' back action from the output
data (make a ``quantum variational measurement'').

Our examples are the following:  
We will begin in Sec.\ \ref{sec:PositionMeasurement} 
with a simple, idealized, instantaneous
single measurement of the position of a single test mass. 
This example will demonstrate that the noise
associated with test-mass quantization and the noise associated with the
meter's quantization are truly independent (though closely
linked), and will illustrate how under some circumstances they can add, 
producing a doubling of the noise power.  Then, in Sec.\
\ref{sec:VonNeumann},  we will analyze the use of a sequence of these
idealized, instantaneous position
measurements to monitor a classical
force that acts on the test mass. This example will illustrate the 
vanishing self-commutator of the output data samples, which arises
from a cancellation of the test-mass-position commutator by the 
measurement-noise commutator; it also will illustrate how signal processing
can remove all influence of test-mass quantization and test-mass 
state reduction from the output data stream.  Our third example
(Sec.\  \ref{sec:PulsedLightMeasurements}) will be a Heisenberg-microscope-like
realization of these instantaneous, idealized position measurements,
in which a pulse of near-monochromatic light is reflected off the
test mass, thereby encoding the test-mass position in a phase shift
of the light.  This example will give reality to the idealized examples
in Secs.\ \ref{sec:PositionMeasurement} and \ref{sec:VonNeumann}, and
will help connect them to the subsequent discussion of interferometric 
gravitational-wave detectors.

In Sec.\ \ref{sec:IFOs} we will use the insights from our pedagogical
examples to prove and elucidate
our three answers [(i), (ii), (iii) above] for gravitational-wave
interferometers, and also for a wide range of other classical force
measurements.  The underpinnings for our answers will be: (a) a proof
that for a quantized electromagnetic wave, such as that entering the final
photodetector of an interferometer, the photon number flux operator commutes
with itself at different times (this flux is the output data
stream), and (b) a proof that all influence of the test-mass quantum
observables 
can be removed from the output data stream by 
appropriate filtering, and for 
conventional
interferometers it is sufficient to
remove all data near the test masses' $\sim 1$ Hz swinging frequency, e.g.\
by
the kind of high-pass
filtering that is routinely used in gravitational-wave detectors.
Our analysis will also elucidate QND interferometer designs based on
(A) squeezed-input states for light and (B) variational-output measurements.

The issues studied in this paper are most efficiently analyzed in
the Heisenberg picture, and the Heisenberg picture gives particularly
clear insights into them.  For this reason, we will use the Heisenberg
picture throughout the body of this paper.  Readers who are uncomfortable
with the Heisenberg picture may find Appendix \ref{app:TripleMeasurement}
reassuring; there we will give a detailed Schroedinger-picture
analysis of the most important of our pedagogical examples, that of
Sec.\  \ref{sec:VonNeumann} 

\section{Pedagogical Examples}
\label{sec:Pedagogy}

\subsection{A single position measurement: \\
``double'' uncertainty relation}
\label{sec:PositionMeasurement}

We begin with a simple pedagogical example of a
single measurement of the position of a single test mass.
The Heisenberg microscope is a famous realization of this example; see Sec.\
\ref{sec:PulsedLightMeasurements}.  

The measurement is idealized as instantaneous and as occuring at time $t=0$.
At times arbitrarily close to $t=0$, the Hamiltonian for the test mass (with
position and momentum $\hat x$ and $\hat p$) and the measuring device (the {\it
meter}, with generalized position $\hat Q$ and generalized momentum $\hat P$)
is
\begin{equation}
H = {\hat p^2\over2\mu} - \delta(t)\hat x \hat P +{\hat P^2\over2M}\;.
\end{equation}
Here $\delta(t)$ is the Dirac delta function, 
$\mu$ is the test mass's mass and $M$ is the generalized mass of the meter.
For pedagogical simplicity we make $M$ arbitrarily large 
so $\hat Q$ and $\hat P$ do
not evolve in the Heisenberg Picture except at the moment of interaction,
and correspondingly we rewrite the Hamiltonian as
\begin{equation}
H = {\hat p^2\over2\mu} - \delta(t)\hat x \hat P\;.
\label{HSimple}
\end{equation}

A simple calculation in the Heisenberg picture gives the following expressions
for the positions and momenta immediately after the measurement, in terms of
those immediately before:
\begin{mathletters}
\begin{eqnarray}
\hat P_{\rm after} &=& \hat P_{\rm before}  \;, 
\label{BeforeAfterA} \\
\hat x_{\rm after} &=& \hat x_{\rm before}  \;, 
\label{BeforeAfterB}\\
\hat Q_{\rm after} &=& \hat Q_{\rm before} - \hat x_{\rm before} \; 
\label{BeforeAfterC}\\
\hat p_{\rm after} &=& \hat p_{\rm before} + \hat P_{\rm before}\;.
\label{BeforeAfterD}
\end{eqnarray}
\label{BeforeAfter}
\end{mathletters}
The meter's generalized position $\hat Q_{\rm after}$
is amplified and read out classically
immediately after the interaction, to determine the test-mass position.
The resulting measured position,  
expressed as an operator,
is $\hat x_{\rm meas} \equiv - \hat Q_{\rm after} = \hat
x_{\rm before} - \hat Q_{\rm before}$ [Eq.\ (\ref{BeforeAfterC})], and 
the measurement leaves the actual
test-mass position operator unperturbed
[Eq.\ (\ref{BeforeAfterB})] but it perturbs the test-mass momentum
[Eq.\ (\ref{BeforeAfterD})].

It is instructive to rewrite Eqs.\ (\ref{BeforeAfterC}) and 
(\ref{BeforeAfterD}) in the form
\begin{mathletters}
\begin{eqnarray}
  \hat x_{\rm meas} & = & \hat x_{\rm before} + \delta \hat x_{\rm meas}\;,
\label{SimpleEqsA}\\
  \hat p_{\rm after} & = & \hat p_{\rm before} + \delta \hat p_{\rm BA}\;,
\label{SimpleEqsB}
\end{eqnarray}
\label{SimpleEqs}
\end{mathletters}
with 
\begin{equation}
\delta \hat x_{\rm meas} = - \hat Q_{\rm before}\;, \quad
\delta \hat p_{\rm BA} = + \hat P_{\rm before}\;.
\label{MeasBADef}
\end{equation} 
The simple equations (\ref{SimpleEqs})
embody the measurement result and its back action; $\hat x_{\rm meas}$ is
the measured value of $\hat x_{\rm before} = \hat x_{\rm after}$, 
$\delta \hat x_{\rm meas}$
is the noise superposed on that measured value by the meter, and $\delta
\hat p_{\rm BA}$ is the back-action impulse given to the test mass by the
meter.  Equations (\ref{SimpleEqs}) are
actually much more general than our simple example; they apply to any
sufficiently quick,\footnote{i.e., quick compared to 
the evolution of the wave function of the
measured quantity, so it can be regarded as constant during
the measurement.
}
``linear'' measurement;
see Eqs.\ (5.2), (5.14) and (5.23) of
Ref.\ \cite{QuantumMeasurement}, 
and see Appendix \ref{app:LinearMeasurements} below.

The initial test-mass position and momentum and the initial meter position
and momentum have the usual commutation relations,
\begin{equation}
[\hat x_{\rm before},\hat p_{\rm before}] = i\hbar = [\hat Q_{\rm before},
\hat P_{\rm before}]\;.
\label{UsualCommutator}
\end{equation}
The second of these and Eqs.\ (\ref{MeasBADef}) imply that the measurement
noise $\delta \hat x_{\rm meas}$ and the 
back-action impulse $\delta \hat p_{\rm BA}$ have this same standard
commutator, but with the sign reversed
\begin{equation}
[\delta\hat x_{\rm meas}, \delta\hat p_{\rm BA}] = -i\hbar\;.
\label{OppositeCommutator}
\end{equation}
This has an important implication:  The measured value of the test-mass
position and the final value of the test-mass momentum commute:
\begin{equation}
[\hat x_{\rm meas},\hat p_{\rm after}] = 0\;.
\label{VanishingCommutator}
\end{equation}
This result, like the simple measurement and back-action 
equations (\ref{SimpleEqs}),
is true not only for this pedagogical example, but also for any other
sufficiently quick, linear measurement; see, e.g., Sec.\ 
\ref{sec:PulsedLightMeasurements} below.  

It is evident from Eqs.\ (\ref{SimpleEqs}) 
and (\ref{MeasBADef})  
that the variances of 
$\hat x_{\rm meas}$ and $\hat p_{\rm after}$ are
influenced by the initial states of both the meter and the test mass:
\begin{eqnarray}
  (\Delta x_{\rm meas})^2 & = & (\Delta x_{\rm before})^2 +
(\Delta Q_{\rm before})^2 \;,
  \label{PositionVariance}\\
  (\Delta p_{\rm after})^2 & = & (\Delta p_{\rm before})^2 +
(\Delta P_{\rm before})^2\;.
\label{MomentumVariance}
\end{eqnarray}
Here we have assumed, as is easy to arrange, that the initial states of the
meter and the test mass are uncorrelated.
Now, the initial states of the test mass and meter are constrained by the
uncertainty relations
\begin{eqnarray}
  \Delta x_{\rm before} \cdot \Delta p_{\rm before} &\ge& \frac{\hbar}{2}\;,
\label{TMUncertainty}\\
  \Delta Q_{\rm before} \cdot \Delta P_{\rm before} &\ge& \frac{\hbar}{2}\;,
\label{MeterUncertainty}
\end{eqnarray}
which follow from the commutators (\ref{UsualCommutator}).
From the viewpoint of
the measurement equations (\ref{SimpleEqs}),
the meter equation (\ref{MeterUncertainty})
is an uncertainty relation
between the noise $\delta \hat x_{\rm meas} = - \hat Q_{\rm before}$ that the
meter superimposes on the
output signal, and the back-action impulse
$\delta \hat p_{\rm BA} = \hat P_{\rm
before}$ that the meter gives to the test mass.
In the Heisenberg microscope,
$\delta \hat x_{\rm meas}$ would be photon shot noise and $\delta \hat p_{\rm
BA}$ would be
radiation-pressure impulse.

The test-mass uncertainty relation (\ref{TMUncertainty}) and meter
uncertainty relation (\ref{MeterUncertainty}) both constrain the product of
the measurement
error (\ref{PositionVariance}) and the final momentum uncertainty
(\ref{MomentumVariance}), and by equal
amounts.  The result is a ``doubling'' of the 
uncertainty relation, so
\begin{equation}
  \Delta x_{\rm meas} \cdot \Delta p_{\rm after} \ge 
2\cdot\frac{\hbar}{2}\;.
\label{DoubledUR}
\end{equation}

This doubling of the uncertainty relation relies crucially on our
assumption that the initial states of the test mass and meter are
uncorrelated.  Correlations can produce a violation of the 
uncertainty relation (\ref{DoubledUR}).  For example, 
initial correlations can be arranged so as to produce (in principle)
a vanishing
total measurement error $\Delta x_{\rm meas} = 0$ and a finite $\Delta p_{\rm
after}$ so the product
$\Delta x_{\rm meas} \cdot \Delta p_{\rm after}$ vanishes --- a result
permitted by the vanishing commutator (\ref{VanishingCommutator}).

\subsection{Monitoring a classical force:\\
``single'' uncertainty relation}
\label{sec:VonNeumann}

As we emphasized in Sec.\ \ref{sec:Questions}, the goal of LIGO-type
detectors is {\it not} to measure any
observables of a test mass, but rather to monitor an external force that
acts on it. Correspondingly, it is desirable to design the measurement so
the output is devoid of any information about the test mass's initial state.
As we shall see, this is readily done in a way that removes
the initial-state information during data processing.  The result is
a ``single'' uncertainty relation: the measurement result is influenced only
by the quantum properties of the meter and not by those of the test mass.

\subsubsection{Von Neumann's thought experiment}
\label{sec:VNThought}

We illustrate this by a variant of a thought experiment devised by
von Neumann \cite{vonneumann} and often used to illustrate issues
in the quantum theory of measurement; see, e.g., \cite{cavesmilburn} and
references therein.
We analyze this thought experiment using the Heisenberg picture in the
body of this paper, and we give a Schroedinger-picture analysis in 
Appendix \ref{app:TripleMeasurement}.  

Our von Neumann thought experiment is a simple generalization of
the position measurement described above.
Specifically,  we consider a
free test mass, with mass $\mu$, position $\hat x$ and momentum $\hat p$, on
which acts a classical force $F(t)$. To monitor $F(t)$, we probe the test
mass instantaneously at times $t=0$, $\tau$, $\ldots$, $(N-1)\tau$ using $N$
independent meters labeled $r=0,1,...,N-1$. 
Each meter is prepared in a carefully chosen state, it
then interacts with the test mass, and then is measured.  We filter the
measurement results to deduce $F(t)$. Meter $r$ has generalized coordinate
and momentum $\hat {Q}_r$ and $\hat {P}_r$, and its free Hamiltonian is
vanishingly small, so $\hat{Q}_r$ and $\hat{P}_r$ do not evolve except at the
moment of interaction. The total Hamilton for test mass plus classical force
plus meters is 
\begin{equation}
  \hat H = {\hat p^2\over 2 \mu} - F(t) \hat x
    - \sum_{r=0}^{N-1} \delta(t-r\tau) \hat x \hat {P}_r\;.
\label{Hamiltonian}
\end{equation}

We denote by $\hat x_0$ and $\hat p_0$ the test-mass position and momentum at
time $t=0$ when the experiment begins, and by $\hat x_r$ and $\hat p_r$ their
values immediately {\it after} interacting with meter $r$, at time $t=r\tau$.
The momentum of meter $r$ is a constant of the motion, so we denote it by
$\hat{P}_r$ at all times. The meter coordinate changes due to the
interaction; we denote its value before the interaction by $\hat
{Q}_r^{\rm before}$ and after the interaction by $\hat {Q}_r$.

It is easy to show, from the Heisenberg equations for the Hamiltonian
(\ref{Hamiltonian}), that the test-mass
position immediately after its $r$'th interaction is

\begin{equation}
  \hat x_r = \hat x_o + {\hat p_o\over \mu}r\tau +
  \sum_{s=0}^{r}\hat{P}_s {(r-s)\tau\over \mu} + \xi_r\;.
  \label{x_r}
\end{equation}
Here the first two terms are the free evolution of the test mass, the third
(with the sum) is the influence of the meters' back-action forces (analog of
radiation-pressure force in an interferometer), and the fourth,
\begin{equation}
  \xi_r \equiv \frac{1}{\mu}\int_0^{r\tau} \int_0^t F(t') dt'dt =
  \frac{1}{\mu}\int_0^{r\tau} (r\tau-t')F(t') dt'\;,
\label{xirDef}
\end{equation}
is the effect of the classical force. The force $F(t)$ is encoded in the
sequence of classical displacements $\{\xi_1,\xi_2, ... , \xi_N\}$. It is
also easy to show from the Heisenberg equations
that the meter's generalized coordinate after interaction
with the test mass is
\begin{eqnarray}
  \hat{Q}_r & = & \hat{Q}_r^{\rm before} - \hat x_r \nonumber\\
  & = & {\hat Q}_r^{\rm before} 
    - \hat x_o - {\hat p_o\over \mu}r\tau - \sum_{s=0}^{r}
    \hat{P}_s {(r-s)\tau\over \mu} - \xi_r\;.
  \label{Qr}
\end{eqnarray}

\subsubsection{Vanishing of the output's self commutator}

The set of final meter coordinates $\vec Q \equiv \{\hat{Q}_0, \hat{Q}_1,
...  ,$ $\hat{Q}_{N-1}\}$ forms the final data string for data analysis. It has
vanishing self commutator,
\begin{equation}
  [\hat{Q}_s, \hat{Q}_r] = 0 \quad \hbox{for all $s$ and $r$}
  \label{CommutatorVN}
\end{equation}
--- a result that can be deduced 
from the
vanishing single-measurement
commutator 
$[\hat x_{\rm meas}, \hat p_{\rm after}] = 0$ [Eq.\
(\ref{VanishingCommutator})] for the earlier of the two measurements.
 
It is instructive to see explicitly how this vanishing commutator arises,
without explicit reference to our single-measurement analysis.
The test-mass contributions to the $Q$'s
[$\hat x_o$ and $\hat p_o$ in Eq.\ (\ref{Qr})] produce
\begin{eqnarray}
[\hat Q_s,\hat Q_r]_{\rm test-mass} &=& 
\left[-x_o-{p_o\over\mu}s\tau,\; -x_o-{p_o\over\mu}r\tau\right] \nonumber\\
&=&\frac{i\hbar (r-s)\tau}{\mu},
\end{eqnarray}
which is the analog of 
Eq.\ (\ref{TMCommutator}) for an interferometer test mass.
This must be cancelled by a contribution
from the meters.  Indeed it is. If (for concreteness) $r>s$, then the
cancelling contribution comes from a commutator of (i) the $\hat{Q}_s^{\rm
before}$
piece of $\hat{Q}_s$ (the noise superposed on the output signal $s$ by meter
$s$) and (ii) the $\hat P_s$ term in $\hat Q_r$ (the noise in the later
measurement produced by the back-action of the earlier measurement):
\begin{eqnarray}
[\hat Q_s,\hat Q_r]_{\rm meter} &=& \left[ \hat Q_s^{\rm before}, - \hat
P_s{(r-s)\tau\over \mu}\right] \nonumber \\
&=& {-i \hbar (r-s)\tau\over\mu}\;.
\end{eqnarray}

In this example,
one can trace these cancellations to the bilinear form $\hat x \hat P_s$ and
$\hat x \hat P_r$ of each piece of the interaction Hamiltonian.  However,
this type of cancellation is far more general than just bilinear
Hamiltonians:  In {\it every sequence of measurements on any kind of system},
by the time a human looks
at the output data stream, its entries have all been amplified to classical
size, and therefore they must all be classical quantities and must commute,
$[\hat Q_s, \hat Q_r] = [Q_s,Q_r] = 0$.  Remarkably, quantum mechanics is so
constructed that, for a wide variety of measurements, the measured values
(regarded as Hermitian observables)
commute even before the amplification to classical size.  
This is true in
the above example. 
It is true in a realistic variant of this example involving pulsed-light
measurements (Sec.\ \ref{sec:PulsedLightMeasurements}). 
It is true in a variant of this example involving continuous measurements
by an electromagnetic wave in an idealized transmission line \cite{gardiner}. 
And, as we shall see in Sec.\ \ref{sec:PhotonFluxCommutator} and 
Appendix \ref{app:VanishingCommutator},
it is
also true for gravitational-wave interferometers --- and indeed for all
measurements
in which the measured results are encoded in the photon number flux of a
(quantized) 
electromagnetic wave; 
i.e., all measurements based on photodetection. 
More generally, it is true for any {\it linear measurement} (Appendix
\ref{app:LinearMeasurements} below, Ref.\ \cite{QuantumMeasurement}, and
Eq.\ (2.34) of Ref.\ \cite{BC3});
and, in fact, all the measurements discussed above, including gravitational-wave
measurements, are linear.

The classical nature of the output signal (the commutation of the data entries)
guarantees that, when a human looks at one data entry, the resulting reduction
of the state of the measured system cannot have any influence on the observed
values of the other data entries.  Correspondingly, we can carry out any
data processing procedures we wish on the $\hat Q_r$, without fear of
introducing new quantum noise.

\subsubsection{Removal of test-mass influence from the output}

Our goal is to measure the classical force
$F(t)$ that acted on the test mass, without any contamination from the
test mass's quantum properties --- more specifically, without any
contamination from uncertainty-principle aspects of the test mass's
initial state.  The initial state {\it does} influence the measured
values $\tilde Q_r$ of the output observables $\hat Q_r$, since
in the Heisenberg Picture the $\hat Q_r$ contain the test mass's
initial position $\hat x_o$ and momentum $\hat p_o$  
[Eq.\ (\ref{Qr})].  Therefore, our
goal translates into finding a data analysis procedure that will
remove from the output data set $\{\tilde Q_1,
\tilde Q_2, \ldots\}$ all influence of the test-mass
initial state (or equivalently all influence of 
$\hat x_o$ and $\hat p_o$), while retaining the influence of $F(t)$.  In
fact, we can do so rather easily, 
regardless of what the test-mass initial state might have been.
As we shall see, our ability to do so relies crucially on the {\it linearity}
of our measurements; in particular, on the fact that the output observables
$\hat Q_r$ are linear in $\hat x_o$ and $\hat p_o$.

To bring out the essence, we shall restrict
ourselves to just three meters, $N=3$. The generalization to large $N$ is
straightforward.

The measured data sample
$\hat Q_r$ is equal to the freely evolving test-mass position
at time $r\tau$,
$\hat x_{\rm free}(t=r \tau ) = \hat x_o + (\hat p_o/\mu)r\tau$
(which is linear in $\hat x_o$, $\hat p_o$), 
plus noise.
Since the free evolution satisfies the equation of motion
$d^2\hat x_{\rm free}/dt^2 = 0$,
it is a reasonable guess that we can remove the influence of $\hat x_o$ and
$\hat p_o$ from the data $\tilde Q_r$ by applying to them the discrete version of a second
time derivative\footnote{In Sec.\ III$\;$C of Ref.\ \cite{caves2}, Caves
uses his path-integral formulation of measurement theory to analyze 
measurements of the discrete second time derivative of the position of a
free particle on which a classical force acts.  His analysis reveals
the same conclusion as we obtain in our pedagogical example:  the 
measured quantity contains information about the force and is devoid of
any influence from the particle's initial state.}
(which is a linear signal processing procedure).
Accordingly, from
the measured values $\{\tilde Q_0, \tilde Q_1, \tilde Q_2\}$ of
$\{\hat Q_0, \hat Q_1, \hat Q_2\}$ in a representative experiment, we construct
the discrete second time derivative
\begin{equation}
\tilde R  = (\tilde Q_2 - \tilde Q_1) - (\tilde Q_1 - \tilde Q_0) =
    \tilde Q_0 - 2\tilde Q_1 + \tilde Q_2 \label{R}\;.
\end{equation}
The following argument shows that all the statistical properties of this
quantity, in a large series of experiments (in which the initial states $|{\rm
in}\rangle$ of the test mass and meters are always the same)
are, indeed, devoid of any influence of $\hat x_o$ and $\hat p_o$,
and thus are unaffected by the test-mass initial 
state.\footnote{The crucial idea of avoiding the 
influence of the test-mass initial state
by monitoring differences of observables 
[$(\hat Q_2 - \hat Q_1) - (\hat Q_1 - \hat Q_0)$ in our case]
is contained in a paper 
and book
by Alter
and Yamamoto\cite{alter_yamamoto,alter_yamamoto_book}.  
Alter and Yamamoto point out that,
for a test mass on which a classical force acts, 
the momentum $\hat p(t)$ at time $t$ and the
momentum $\hat p(0)$ at time 0 are correlated in that
$\hat p(t) = \hat p(0) + \int_0^t dt' F(t')$; so, if one measures
$\hat p(t) - \hat p(0) = \int_0^t dt' F(t')$, one thereby can
get information about the force without any contaminating influence
of the test-mass initial state.  
They say (page 96 of \cite{alter_yamamoto_book}) that this is so not only when
one measures directly the difference $\hat p(t) - \hat p(0)$ (as in 
Sec.\ 7.2.2 of their \cite{alter_yamamoto_book}), but also when the difference
is determined computationally from the results of measurements of
$\hat p(t)$ and $\hat p(0)$ [an analog of our way of monitoring
$(\hat Q_2 - \hat Q_1) - (\hat Q_1 - \hat Q_0)$].
When going on to discuss 
position measurements, Alter and Yamamoto note that
$\hat x(t) - \hat x(0) = \hat p(0) t/m + \int_0^t dt' \int_0^{t'} dt'' 
F(t'')/m$, so a measurement of $\hat x(t) - \hat x(0)$ {\it is}
contaminated [via $\hat p(0)t/m$]
by noise from the test-mass initial state.  Examining
this contamination, they conclude that ``force detection via position
monitoring of a free mass is limited by ... the SQL''
\cite{alter_yamamoto}.  
While this
conclusion is correct when one monitors $\hat x(t) - \hat x(0)$
in the manner envisioned by Alter and Yamamoto,
it is incorrect for the alternative strategy embodied in our model
problem.  Instead of monitoring $\hat x(t) - \hat x(0)$, one should
monitor $\hat x(0) - 2 \hat x(t) + \hat x(2t)$, which for a free mass
is independent of both $\hat x_o \equiv \hat x(0)$ and $\hat p_o \equiv
\hat p(0)$.  Then the measurement output contains information about 
about the force $F(t)$, uncontaminated by any influence of the test-mass
initial state.
}

These statistical properties are embodied in the means, over all the
experiments, of arbitrary functions $G(\tilde R)$.
The theory of measurement tells us that, because the $\hat Q$'s all commute,
the computed mean of $G(\tilde R)$ is given by
\begin{equation}
[\hbox{computed mean of } G(\tilde R)] =
\langle {\rm in} | G( \hat R) | {\rm in} \rangle\;,
\label{GExpectation}
\end{equation}
where $\hat R$ is the operator corresponding to $\tilde R$
\begin{eqnarray}
\hat R  &=&
    \hat Q_0 - 2\hat Q_1 + \hat Q_2 \
= - ( \xi_0 -2\xi_1 + \xi_2 ) + \nonumber\\
&&\quad + \left[
    \hat Q_0^{\rm before}-2\hat Q_1^{\rm before}-{\hat P_1\tau\over \mu}+\hat
Q_2^{\rm before}
  \right]\; \label{hatR}
\end{eqnarray}
cf.\ Eq.\ (\ref{Qr}).
Because $\hat R$ is independent of $\hat x_o$ and $\hat p_o$,
{\it the computed
mean (\ref{GExpectation}) and thence all the measurement statistics of
$\tilde R$
will be completely independent of the test-mass
quantum mechanics, and in particular independent of the test mass's
initial state.}  Moreover, Eq.\ (\ref{GExpectation}) implies that, so far as
measurement results and statistics are concerned, measuring the $\hat Q$'s and
then computing $\tilde R$ is completely equivalent to measuring $\hat R$
directly.

Although $\hat R$ is independent of $\hat x_o$ and $\hat p_o$
it contains
\begin{equation}
  \xi_0 - 2\xi_1 + \xi_2 = \frac{1}{\mu}\int_0^{2\tau}(\tau-|t-\tau|)F(t)dt
\equiv {\tau^2\over \mu}\bar F\;,
\label{xiVal}
\end{equation}
where $\bar F$ is a weighted mean of the classical force $F$ over the time
interval
$0<t<2\tau$;
cf.\ Eq.\ (\ref{xirDef}).\footnote{Notice that, 
aside from meter noise, $\xi_r$ is equal to $\hat x(t_r) - \hat p(0) t_r /\mu$
[Eq.\ (\ref{x_r})], which is a QND observable 
(as M.B.\ Mensky pointed out long ago). Therefore, the
quantity $\hat R$ that we measure can be regarded as a discrete second
time derivative of a QND observable --- which suggests that it can be
the foundation for a QND measurement; see Sec.\ \ref{sec:BeatSQL} below.}
Thus, {\it this measurement of $\hat R$ is actually a measurement of
$\bar F$, and is contaminated by quantum noise from the meters but {\bf not}
by quantum noise from the test mass.}  The only role of the quantum mechanical
test mass is to feed the 
classical signal $\bar F$ and the meter back-action noise $\hat
P_1 \tau/m$ into the output.

For those readers who are uncomfortable with our use of the Heisenberg
picture to derive this very important result, we present a Schroedinger-picture
derivation in Appendix \ref{app:TripleMeasurement}.

This three-meter thought experiment 
is a prototype for our  discussion
of gravitational-wave interferometers in Sec.\ \ref{sec:Filter}.
There as here, the {\it linearity of the output} 
in the test-mass initial
positions and momenta will enable us to find
a linear signal processing procedure that removes the initial-state influence. 
Here that procedure was a discrete second time derivative.  For an 
interferometer it will be a discrete Fourier transform of the measured
photon flux (the output), and a discarding of Fourier components at the 
test masses'
natural frequencies (the 1 Hz pendular swinging frequency in the case
of conventional interferometers).  

For an elegant path-integral analysis
of the removal of test-mass initial conditions from the output of
measurements of any harmonic oscillator on which a classical force acts,
see the last portion of Sec.\ III$\;$C of Caves \cite{caves2}.

\subsubsection{The SQL for the classical-force measurement}
\label{ClassicalSQL}

How small can the test-mass noise be?  A ``naive'' optimization of the
meters leads to the standard quantum limit on the measured force,
in the same way as a ``naive'' optimization of a
gravitational-wave interferometer's design (forcing it to retain
the conventional LIGO-I optical topology but optimizing its laser power)
leads to the gravitational-wave SQL.  Specifically:

Let the three meters all be prepared in initial states that are ``naive''
in the sense that they have
no correlations between their coordinates and momenta. Then
Eqs.\ (\ref{hatR}) and (\ref{xiVal}) imply that the variance of the
measured mean force is
\begin{eqnarray}
(\Delta \bar F)^2 &=& {\mu^2\over \tau^4}
\Big[ (\Delta Q_0^{\rm before})^2
+ (2\Delta Q_1^{\rm before})^2
+ \left({\Delta P_1 \tau\over\mu}\right)^2  \nonumber\\
&& + (\Delta Q_2^{\rm before})^2 \Big]\;.
\nonumber\\
\label{DeltabarF}
\end{eqnarray}
Obviously, this variance is minimized by putting meters 0 and 2 into (near)
eigenstates of their coordinates, so $\Delta Q_0^{\rm before} = \Delta Q_2^{\rm
before} = 0$.  To minimize the noise from meter 1, we require that it
have the smallest variances compatible with
its uncertainty relation,
\begin{equation}
  \Delta Q_{1}^{\rm before}\Delta P_{1} = \frac{\hbar}{2} \; ,
\end{equation}
and we adjust the ratio $\Delta Q_1^{\rm before}/\Delta P_1$ so as to minimize
$(\Delta \bar F)^2$. The result is
\begin{equation}
(\Delta \bar F)^2 = {2\mu \hbar \over \tau^3}\;,
\label{barFSQL}
\end{equation}
which is the SQL for measuring a classical force, up to a factor of order unity;
cf.\ Sec.\ 8.1 of Ref.\ \cite{QuantumMeasurement}.

It is evident from this analysis that {\em the true physical origin of the
SQL in classical force measurements is the meter's noise}, not the test-mass
noise. On the other hand,
the quantum
properties of the meter and of the test mass are intimately coupled
through the requirement that the meter commutators cancel the test-mass
commutator in the measurement output, so that $[\hat Q_r, \hat Q_s] = 0$
[Eq.\ (\ref{CommutatorVN})]. This intimate coupling --- which, as we have
discussed, has enormous generality --- ensures that
the SQL can be derived equally well from test-mass considerations and
from meter considerations. We saw this explicitly in
Sec.\  \ref{sec:Questions} for
an interferometric gravitational-wave detector.

\subsubsection{Beating the SQL}
\label{sec:BeatSQL}

Equation (\ref{hatR}) suggests a way to beat the classical-force SQL
and, in fact, achieve arbitrarily high accuracy:  As in our ``naive''
optimization, before the measurement we place
meters 0 and 2 in
(near) eigenstates of their coordinates,
so $\Delta Q_0 = \Delta Q_2 = 0$,
but instead of putting meter 1 in a ``naive'' state with uncorrelated
coordinate and momentum, we place it in a
(near) eigenstate of 
\begin{equation}
\hat Q_1^{\rm squeeze} \equiv \hat Q_1^{\rm before}-\hat P_1\tau/2\mu\;.
\label{IdealSqueezed}
\end{equation}
(This meter-1 state is analogous to the squeezed-vacuum state, which
Unruh \cite{unruh} has proposed be inserted into a conventional
interferometer's dark port in order to beat the gravitational-wave SQL;
see Sec.\ \ref{sec:PulsedLightMeasurements} below.)
These initial meter states,
together with Eqs.\ (\ref{hatR})
and (\ref{GExpectation}),
guarantee that the variance of the computed quantity $\tilde R$ vanishes
$\Delta \tilde R = 0$, and thence [via Eqs.\  (\ref{xiVal}) and
(\ref{hatR})] that
the variance of the measured mean force vanishes, $\Delta \bar F = 0$.
Thus, by putting the initial state of meter 1 into the analog of a squeezed
vacuum state, we can achieve an arbitrarily accurate measurement of
$\bar F$.

The SQL can also be evaded by modifying the meters' measured quantitites
instead of modifying their initial states.  Specifically,
measure $\hat Q_0$ and $\hat
Q_2$ as before, but on meter 1 instead of measuring the coordinate $\hat Q_1$,
measure the following linear combination of the coordinate and
momentum (with the coefficient $\alpha$ to be chosen below):
\begin{eqnarray}
  \hat Q^{\rm var}_1 &=& \hat Q_1+\alpha \hat P_1 \nonumber \\
    & = &
    \hat Q_1^{\rm before} - \hat x_0 - \frac{\hat p_0}{\mu}\,\tau -
    \frac{\hat P_0}{\mu}\tau - \alpha \hat P_1 - \xi_1\;.
\label{calQ1}
\end{eqnarray}
From Eqs.\ (\ref{calQ1}), (\ref{Qr}) and (\ref{CommutatorVN}), we see that
the output observables $\{\hat Q_0,\hat Q^{\rm var}_1, \hat Q_2\}$ all commute
with each other.  Therefore, when we combine their measured values
into the discrete second time derivative
\begin{equation}
\tilde R_{\rm var} \equiv \tilde Q_0 - 2 \tilde  Q^{\rm var}_1 +
\tilde Q_2\;,
\end{equation}
its statistics will be the same as if we had directly measured the
corresponding operator
\begin{eqnarray}
   & &\hat R_{\rm var} =  \tilde Q_0 - 2 \tilde  Q^{\rm var}_1 +
\tilde Q_2 =
-(\xi_0-2\xi_1+\xi_2) \nonumber \\
     & & \quad +
     \left[
       \hat Q_0^{\rm before} - 2\hat Q_1^{\rm before} +
       \frac{\hat P_1}{\mu}\tau - 2\alpha \hat P_1 + \hat Q_{2}^{\rm before}
     \right]\;.
\label{hatRVar}
\end{eqnarray}
Evidently, we should choose $2\alpha=\tau/\mu$,
so the quantity measured is
\begin{equation}
\hat Q^{\rm var}_1 = \hat Q_1 +{\hat P_1}{\tau\over2\mu}\;.
\label{IdealVar}
\end{equation} 
Then Eqs.\ (\ref{hatRVar})
and (\ref{xiVal}) imply that
\begin{equation}
\hat R_{\rm var} = - {\tau^2\over \mu}\bar F + \hat Q_0^{\rm before} - 2
\hat Q_1^{\rm before} + \hat Q_2^{\rm before}\;.
\label{Rvar1}
\end{equation}
Therefore,
{\em by measuring our chosen linear combination of meter 1's coordinate and
momentum, and then computing the discrete second time derivative,
we have succeeded in removing from our output observable
$\hat R_{\rm var}$
not only the test-mass
variables $\hat x_o$, $\hat p_o$, but also the back-action influence of the
meters on the measurement (all three $\hat P_r$'s)!} Correspondingly,
by putting the meters
into ``naive'' initial states (states with no position-momentum
correlations) that are near eigenstates of their coordinates
(so $\Delta Q_0$, $\Delta Q_1$, $\Delta Q_2$ are arbitrarily small and
the back-action fluctuations
$\Delta P_0$, $\Delta P_1$, $\Delta P_2$ are arbitrarily large), then from
the computed quantity $\tilde R_{\rm var}$, we can infer the mean
position $\bar F$ with arbitrarily good precision.

This strategy was devised, in the context of optical measurements of test
masses, by Vyatchanin, Matsko and Zubova \cite{VMZ,VM1,VM2,VL}, and is called
a {\it Quantum Variational Measurement}.  A gravitational-wave interferometer
that utilizes it (and can beat the SQL) 
is called a {\it Variational Output Interferometer}
\cite{kimble}.

Of course, one can also beat the SQL for force measurements by a combination
of putting the meters into initially squeezed states and performing a quantum
variational measurement on their outputs.  A gravitational-wave detector
based on this mixed strategy is called a {\it Squeezed Variational
Interferometer}, and may have practical advantages over
squeezed-input and variational-output interferometers \cite{kimble}.

\subsection{Pulsed-light measurements of test-mass position}
\label{sec:PulsedLightMeasurements}

Our two pedagogical examples (single position measurement,
Sec.\ \ref{sec:PositionMeasurement}, and classical force measurement, Sec.\
\ref{sec:VonNeumann}) can be realized using pulsed-light measurements of the
test-mass position.  We exhibit this realization in part to lend reality to
our highly idealized examples, and in part as a bridge from
those simple examples to
gravitational-wave interferometers with their far greater complexity
(Sec.\ \ref{sec:IFOs} below).

In each pulsed-light measurement we reflect a laser light pulse, with
carrier frequency $\omega_o$ and Gaussian-profile duration $\tau_o$, 
off a mirror on the front face of the test mass, and from the light's 
phase change we deduce the test-mass position $\hat x$ averaged over the pulse. 
This is a concrete realization not only of the pulsed measurements of 
our pedagogical examples, but also of a Heisenberg microscope.  We
presume that the pulse duration $\tau_o$ is long compared to the light's 
period $2\pi/\omega_o$, but short compared to the time $\tau$ between
measurements.  

We shall analyze in detail one such pulsed measurement.  The electric field
of the reflected wave, at some fiducial location, is 
\begin{eqnarray}
&& \hat E(t) = \sqrt{\frac{2\pi\,\hbar \omega_0}{cS} }\Bigg( 
	e^{-i\omega_0 t}\times \nonumber \\
&&\times \left[ A_0 e^{-t^2/2\tau_0^2}\left(
	1+\frac{2i\omega_0}{c}\ \hat x(t)\right) +  \hat a(t)  \right]
	+\mbox{h.c.}\Bigg),
\label{ElectricField}
\end{eqnarray}
where $A_0$ is the pulse's amplitude, $S$ is its cross sectional area, 
$c$ is the speed of light, $2(\omega_0/c)\hat x(t)$ is the phase shift
induced by the test-mass displacement $\hat x(t)$, ``h.c.'' means Hermitian
conjugate, and 
$\hat a(t)$ is the electric field's amplitude operator.
Because  we are concerned only about timescales of order the pulse duration 
$\tau_0$ or longer, which means side-band frequencies $\alt 1/\tau_0 \ll \omega_0$,
we can use the {\it quasimonochromatic} approximation to the
commutation relation for $\hat a(t)$ \cite{gardiner1}:
\begin{equation}
\left[ \hat a(t),\hat a^\dag(t') \right] = \delta(t-t')\;.
\end{equation}
Note that, when decomposed into quadratures with respect to the carrier
frequency, this electric field is
\begin{equation}
\hat E(t) = \hat E_A(t) \cos\omega_o t + \hat E_\phi(t) \sin\omega_o t\;,
\label{EDecompose}
\end{equation}
where $\hat E_A$ and $\hat E_\phi$, the amplitude and phase 
quadratures
(i.e., the quadrature components oriented along
and perpendicular to the amplitude direction in the quadrature plane)
are given by
\begin{mathletters}
\begin{eqnarray} 
\hat E_A &=& 2\sqrt{2\pi \hbar\omega_o\over cS}\left[ A_o e^{-t^2/2\tau_o^2} 
+ \left({\hat a(t) + \hat a^\dag(t)}\over 2\right) \right]\;, \label{EA} \\
\hat E_\phi &=& 2\sqrt{2\pi \hbar\omega_o\over cS}\Big[
2A_o {\omega_o\over c}e^{-t^2/2\tau_o^2} \hat x(t) \nonumber \\
&&\quad\quad\quad\quad\quad + 
\left({\hat a(t) - \hat a^\dag (t)}\over {2i}\right) \Big]\;. \label{Ephi}
\end{eqnarray}
\label{EQuadratures}
\end{mathletters} 

The power $\hat W(t)$ in the incident wave can be written as the sum of 
a mean power $\langle W(t)\rangle$ and a fluctuating (noise) 
part $\tilde W(t)$:
\begin{mathletters}
\begin{eqnarray}
\hat W(t)&=& S c \frac{\overline{\hat E^2(t)}}{4\pi}
=\langle W(t)\rangle + \tilde W(t), \\
\langle W(t)\rangle &=& \hbar \omega_0\, A_0^2\,e^{-t^2/\tau_0^2},\\
\tilde W(t) & = & 2\hbar \omega_0\, A_0\,e^{-t^2/2\tau_0^2} \left(
	\frac{\hat a(t)+ \hat a^\dag(t)}{2} \right)\;.
\end{eqnarray}
\end{mathletters}
Here the over bar means ``average over the carrier period''.
The light-pressure force on the mirror is $\hat F(t) = 2 \hat W(t)/c$.  
The fluctuating part of this, $\tilde F(t) = 
2\tilde W(t)/c$, is the back-action of the measurement on the test mass,
and it produces the back-action momentum change  
\begin{eqnarray}
\delta \hat p_{\rm BA}
&=&\int_{-\infty}^\infty dt\, \frac{2\tilde W(t)}{c}= \nonumber \\
&=&	\frac{4\hbar \omega_0}{c}\, A_0\,
	\int_{-\infty}^\infty dt\,e^{-t^2/2\tau_0^2}\left( 
	\frac{\hat a(t)+ \hat a^\dag(t)}{2}\right)\;.
\label{deltapBA}
\end{eqnarray}
The test-mass momentum before and after the pulsed measurement are related
by 
\begin{equation}
\hat p_{\rm after} = \hat p_{\rm before} + \delta \hat p_{\rm BA}\;.
\label{pafterP}
\end{equation}

The experimenter deduces the phase shift $(2\omega_o/c)\hat x(t)$ and
thence the test-mass displacement $\hat x(t)$ by measuring the 
electric field's phase quadrature $\hat E_\phi$ (e.g., via interferometry or
homodyne detection).
More precisely, the experimenter measures the phase quadrature 
integrated over the pulse, obtaining a result proportional to
\begin{eqnarray}
\hat x_{\rm meas} &=& \sqrt{cS\over 2\hbar\omega_o} 
{c\over 4 \pi\omega_0\tau_0 A_0}\int_{-\infty}^{+\infty}
e^{-t^2/2\tau_0^2} \hat E_\phi(t) dt \nonumber\\
&=& \hat x + \delta \hat x_{\rm meas}\;;
\label{xafterQ}
\end{eqnarray}
cf.\ Eq.\ (\ref{Ephi}).
Here $\hat x$ is the mirror position averaged over the short pulse,
$\hat x_{\rm meas}$ is the measured value of $\hat x$, and
$\delta \hat x_{\rm meas}$ is the measurement noise superposed on the
output by the light pulse
\begin{equation}
\delta \hat x_{\rm meas} 
= \frac{c}{2\sqrt{\pi}\, \omega_0\tau_0\, A_0}
	\int_{-\infty}^\infty dt\, e^{-t^2/2\tau_0^2}\,	
	\left(\frac{\hat a(t)- \hat a^\dag(t)}{2i}\right)\;.
\label{deltaxmeas}
\end{equation}
It is straightforward, from the commutator $[\hat a(t),\hat a^\dag(t')] =
\delta(t-t')$, to show that the measurement noise and the back-action
impulse have the same commutator
\begin{equation}
\left[\delta \hat x_{\rm meas},\delta \hat p_{\rm BA}\right] = -i \hbar
\label{SameCommutator}
\end{equation}
as for the idealized single measurement of Sec.\ \ref{sec:PositionMeasurement}
[Eq.\ (\ref{OppositeCommutator})], and correspondingly the mirror's measured
position and its final momentum commute,
\begin{equation}
[\hat x_{\rm meas}, \hat p_{\rm after}] = 0\;.
\label{FinalCommutator}
\end{equation} 

The fundamental equations 
(\ref{xafterQ}), (\ref{pafterP}), (\ref{SameCommutator}) and
(\ref{FinalCommutator})
for this pulsed-light measurement are the
same as those 
(\ref{SimpleEqs}), (\ref{OppositeCommutator}),
(\ref{VanishingCommutator})
for our idealized single measurement, and this
measurement is thus a realistic variant of the idealized one.  Similarly,
a sequence of pulsed-light measurements can be used to monitor a classical
force acting on a mirror, and the fundamental equations for such
measurements are the same as for the idealized example of Sec.\ 
\ref{sec:VonNeumann}.

In such pulsed-light experiments, the measurement noise $\delta
\hat x_{\rm meas}$  
is proportional to the fluctuations of the light's 
phase quadrature 
$\hat E_\phi$ [Eqs.\ (\ref{Ephi}) and (\ref{deltaxmeas})],
and the back-action impulse $\delta\hat p_{\rm BA}$ is 
proportional to the fluctuations of its amplitude quadrature 
$\hat E_A$ [Eqs.\ (\ref{EA}) and (\ref{deltapBA})].
Of course, experimenters can measure any quadrature
of the reflected light pulse that they wish.  To achieve a QND {\em
quantum variational} measurement of a classical force acting on the
test mass \cite{VMZ,VM1,VM2,VL}, the experimenter should measure
$\hat Q^{\rm var}_1 = \hat Q_1 + \hat P_1\tau/2\mu$ in the language of our 
idealized thought experiment [Eq.\ (\ref{IdealVar})], which [by Eqs.\
(\ref{MeasBADef})] translates into
$-\delta \hat x_{\rm meas} + \delta \hat p_{\rm BA}\tau/2\mu$ 
plus the light's signal and carrier, which in turn is a 
specific linear combination of the light's amplitude and phase
quadratures $\hat E_A$ and $\hat  E_\phi$
[Eqs.\ (\ref{EQuadratures}), (\ref{deltaxmeas}), (\ref{deltapBA})].
The experimenter can also 
prepare the incident pulse in a {\em squeezed state}, 
in the manner required for an Unruh-type \cite{unruh} QND measurement
of the classical force.  In the language of our idealized thought 
experiment, the desired squeezed state is a (near) eigenstate of
$\hat Q_1^{\rm squeeze} = \hat Q_1 - \hat P_1\tau/2\mu$ 
[Eq.\ (\ref{IdealSqueezed})], which translates into a near eigenstate
of $\delta \hat x_{\rm meas} + \delta \hat p_{\rm BA}\tau/2\mu$
[cf.\ Eqs.\ (\ref{MeasBADef})], or equivalently a near eigenstate of
a specific linear combination of $\hat E_A$ and $\hat E_\phi$.

\section{Gravitational-Wave Interferometers and Other Photodetection-Based
Devices}
\label{sec:IFOs}

We now turn our attention to gravitational-wave interferometers and other
real, high-precision devices for monitoring classical forces that act on
test masses.  Our goal is to prove that for these devices, as for our
idealized examples, the force-measurement precision can be made completely
independent of the test mass's quantum properties, 
including its initial state
and that this can be achieved by an appropriate filtering of the output
data stream.

As in our examples, this conclusion relies on the vanishing
commutator of the 
observables that constitute 
the output data stream.
We shall now discuss the nature of the output data stream and show
that its commutator does, indeed, vanish.

\subsection{Vanishing commutator of the output}
\label{sec:PhotonFluxCommutator}

For interferometers and many other force-monitoring devices, the data
stream, shortly before amplification to classical size, is encoded
in an output light beam, and that beam is sent into a photodetector which
monitors its photon number flux $\hat {\cal N}(t)$.
The photodetector and associated
electronics integrate up $\hat {\cal N}(t)$ over time intervals with duration
$\tau$ long compared to the light beam's carrier period, $\tau \gg
2\pi/\omega_o \sim 10^{-15}$ s, but short compared to the shortest timescales
on which the classical force changes ($\tau \ll \tau_{\rm GW}
\sim 10^{-3}$ s for the gravitational waves sought by interferometers).
For LIGO-I interferometers, the integration time has been chosen to be
$\tau = 5\times
10^{-5}$ s.
The result is a discretized output data stream, whose Hermitian
observables are the numbers of photons in the successive data samples,
\begin{equation}
\hat N_j = \int_{-\infty}^\infty s(t-t_j) \hat{\cal N}(t) dt\;.
\label{Nj}
\end{equation}
Here $t_j = j \tau_0$ is the time of sample $j$, and $s(t)$ is a sampling
function approximately equal to unity during a time interval $\Delta t =
\tau_0$ centered on $t_j$ and zero outside that time interval.

The photon number samples $\hat N_j$ are the analogs, for an interferometer
or other force-monitoring device, of the meter coordinates $\hat Q_j$ in the
idealized example of Sec.\ \ref{sec:VonNeumann}.

In Appendix \ref{app:VanishingCommutator} we show that {\it for any 
free 
light beam,
the number flux operator, evaluated at a fixed plane orthogonal to the
optic axis (e.g.\ at the entrance to the photodetector) self commutes},
\begin{equation}
[\hat {\cal N}(t), \hat {\cal N}(t')] = 0\;.
\label{calNcommute}
\end{equation}
This guarantees, in turn, that all the output photon-number data samples
(\ref{Nj}) commute with each other
\begin{equation}
[\hat N_j, \hat N_k] = 0\;.
\label{NjCommute}
\end{equation}
As we shall see below [Eq.\ (\ref{NTestMass})], the initial position and
momentum of the test mass, $\hat x_o$ and $\hat p_o$, appear linearly in
the output variables $\hat{\cal N}(t)$ and $\hat N_j$.  They obviously will
produce nonzero contributions to the output commutators.  As in our
simple examples
(Sec.\ \ref{sec:Pedagogy}), these nonzero test-mass contributions must be
cancelled by
identical nonzero contributions from noncommutation of the measurement noise
(photon shot noise) and the back-action noise (radiation-pressure noise).

\subsection{Devising a filter to remove test-mass quantum noise} 
\label{sec:Filter}

The vanishing output commutators constitute
our first underpinning for freeing the measurements from the influence
of test-mass quantization.  As in the idealized measurements of Sec.\
\ref{sec:VonNeumann}, the vanishing commutators guarantee
a key property of the data analysis:
If, from each specific realization of the output data stream
$\{\tilde N_1, \tilde
N_2, \ldots \}$, our data analysis produces a new set of quantities (the
``filtered output variables'')
\begin{equation}
\tilde R_J (\tilde N_1, \tilde N_2, \ldots)\;,
\label{tildeRN}
\end{equation}
then the statistics of these $\tilde R_J$ will be identically the same as
if we had directly measured the corresponding observables
\begin{equation}
\hat R_J (\hat N_1, \hat N_2, \ldots)\;,
\label{hatRN}
\end{equation}
rather than computing them from the measured $\tilde N_j$'s.  Therefore,
we can regard our interferometer (or other device) as measuring
the filtered output observables $\{\hat R_1, \hat R_2, \ldots\}$, whatever
those observables may be.

By analyzing the test-mass dynamics of the interferometer (or other
measuring device) in the Heisenberg picture, one can learn how the 
test-mass initial position
$\hat x_o$ and momentum $\hat p_o$ influence the operators $\{\hat N_1, \hat
N_2, \ldots\}$. One can then deduce a set of filtered observables
$\{\hat R_1, \hat R_2, \ldots\}$ in which $\hat x_o$ and $\hat p_o$ do
not appear but the gravitational-wave or other classical force
information is retained. (These will be the
analogs of $\hat R = \hat Q_0-2\hat Q_1+\hat Q_2$ [Eq.\ 
(\ref{hatR})] in our simple model problem).  {\it The filter that
leads from $\{\hat N_1, \hat
N_2, \ldots\}$ to $\{\hat R_1, \hat R_2, \ldots\}$, when applied to the
output (c-number) data $\{\tilde N_1, \tilde    
N_2, \ldots\}$ to produce $\{\tilde R_1, \tilde R_2, \ldots\}$, is guaranteed 
to remove all influence of $\hat x_o$ and $\hat p_o$, and thence all
influence of the test-mass initial state.}

\subsubsection{Influence of $\hat x_o$ and $\hat p_o$ on the output data} 

To make this more specific,
let us explore how $\hat x_o$ and $\hat p_o$ influence the output data train.

To very high accuracy (sufficient for our purposes), interferometers (and most
other force-measuring devices) are {\it linear}.  The inputs are: (i)
the test-mass
position $\hat x(t)$ [actually, the difference between 
four test-mass positions in
the case of an interferometer; Eq.\ (\ref{xDef})], and (ii) the electric field
operators $\hat E_a(t)$, $a=1,2,\ldots$ for the field fluctuations that
enter the interferometer at the bright port, at the dark port, and at all
light-dissipation locations (e.g., at mirrors where bits of light scatter out
of the optical train and reciprocally new bits of field fluctuations scatter
into it); see, e.g., the detailed analysis of interferometers in Ref.\
\cite{kimble}.  The output photon flux is a linear functional of these inputs,
\begin{equation}
\hat {\cal N}(t)
= \int_{-\infty}^t \left[ K_x(t-t') \hat x(t') + \sum_a K_a(t-t')
\hat  E_a(t') \right] dt'\;;
\label{LinearOutput}
\end{equation}
cf.\ the discussion in Appendix \ref{app:VanishingCommutator}.  The $\hat
 E_a$ terms constitute the photon shot noise (analogs of $\hat Q_r^{\rm
before}$ in our idealized example, Sec.\ \ref{sec:VonNeumann}).

The test-mass initial observables $\hat x_o$ 
and $\hat p_o$ enter $\hat {\cal N}(t)$ and thence $\{\hat N_1, \hat N_2, 
\ldots\}$, through $\hat x(t)$
in a manner governed by the test masses' free
dynamics.  The nature of that free dynamics depends on the interferometer
design.  We shall consider two examples in turn: interferometers with
pendular dynamics, and signal-recycled interferometers.  These examples
should be easily extendable to any other type of interferometer than
might be conceived in the future.

\subsubsection{Interferometers with pendular dynamics}

In
conventional gravitational-wave interferometers (e.g.\ LIGO-I, VIRGO
and TAMA)
and in the QND interferometers analyzed by Kimble et.\ al.\
\cite{kimble}, the test masses swing sinusoidally at $\sim 1$ Hz
frequency in response to their suspensions' pendular restoring force
(as modified slightly by the optical cavities' radiation-pressure force):
\begin{equation}
\hat x_{\rm free}(t) = \hat x_o \cos \omega_m t + {\hat p_o\over
\mu\omega_m}\sin\omega_m t\;.
\label{xFree}
\end{equation}
Here $\mu$ is the reduced mass (1/4 the actual
mass of one test mass in the case of an interferometer) and $\omega_m \sim 2\pi
\times 1$~Hz is the pendular swinging frequency.  
There is no significant damping of the free motion 
(3.7)
because the experimenters take great
pains to liberate the test masses from all damping; the typical damping times
in LIGO-I are of order a day, and 
in advanced interferometers (LIGO-II and beyond) will be
of order a year or more \cite{lsc,damping}, which is far longer than the
data segments used in the data analysis.

Superimposed on the
free test-mass dynamics (\ref{xFree}) are (i) the influence $\xi_{\rm GW}(t)$
of the gravitational-wave signal, (ii) the ``back-action'' influence
$\hat x_{\rm BA}(t)$ of
the light's fluctuating radiation pressure (which is linear in
the input fields $\hat E_a$ and is the
analog of the $\hat P_r$ and $\delta p_{\rm BA}$ 
of our discrete model problems), and (iii) the influence
$\xi_{\rm other}(t)$ of  
a variety of other forces --- 
low-frequency
feedback forces from servo systems, thermal-noise forces,
seismic vibration forces, etc: 
\begin{equation}
\hat x(t) = \hat x_{\rm free}(t) 
 +  \xi_{\rm GW}(t) + \hat x_{\rm BA}(t) +  \xi_{\rm other}(t)\;.
\label{xTrue}
\end{equation}

Inserting Eq. (\ref{xFree}) into
(\ref{xTrue}) and then (\ref{xTrue}) into
(\ref{LinearOutput}) we see that, for a test-mass with pendular dynamics,
the initial test-mass position and
momentum operators appear in the output flux operator in the form
\begin{eqnarray}
\hat {\cal N}(t) &=& \int_{-\infty}^t K_x(t-t') \left[
\hat x_o \cos\omega_m t' + {\hat p_o\over\mu\omega_m}\sin\omega_m t'
\right] dt' \nonumber\\
&& + \hbox{(other contributions)}\;.
\label{NTestMass}
\end{eqnarray}
The interferometer's transfer function $K_x(t-t')$ is 
independent of
absolute time and thus transforms frequency-$\omega_m$ inputs into
frequency-$\omega_m$ outputs.  Therefore, $\hat x_o$ and $\hat p_o$ appear
in the output solely at frequency $\omega_m/2\pi \sim$~1 Hz.
Now, because the output data generally have large noise
(seismic and other) at frequencies below $\sim 10$ Hz, it is routine, in
interferometers, to high-pass filter the output data so as to remove
frequencies below $\sim 10$ Hz.  When one does so,
{\em one automatically removes all influence of $\hat x_o$ and $\hat p_o$
from the filtered data $\tilde R_J$} [Eq.\ (\ref{tildeRN})].  
This is a precise analog of
applying the discrete second time derivative to the output data in our
simple example (Sec.\ \ref{sec:VonNeumann}) so as to remove
$\hat x_o$ and $\hat p_o$ from the data; 
and it is a realization of a
general class of measurement procedures, for a harmonic oscillator on
which a classical force acts, that is analyzed by Caves using his
path integral formalism (last part of Sec.\ III$\;$C of Ref. \cite{caves2}).

\subsubsection{Signal-recycled interferometers}

A signal-recycling mirror, placed at an interferometer's output
port, sends information about the test-mass position $\hat x(t)$  back into the
interferometer as part of the back-action (radiation-pressure) force,
and thereby alters the free test-mass dynamics.  The altered free dynamics
have been analyzed in detail by Buonanno and Chen \cite{BC3}; they find
that the test masses and the interferometer's side-band light form a 
coupled system with four degrees of freedom, so $\hat x_o$
and $\hat p_o$ appear in $\hat x_{\rm free}(t)$, and thence in $\hat x(t)$
and thence in $\hat {\cal N}(t)$ at four discrete frequencies
$\omega_A$ ($A=1,2,3,4$).  Correspondingly, in the output data train,
the influence of the test-mass initial state is confined to the
Fourier components at the frequencies $\omega_A$.  

If these frequencies were real, then one could remove the influence
of the test-mass
initial state from the data by filtering out the data's Fourier
components at these four frequencies.  However, as Buonanno and
Chen \cite{BC3} discuss, such filtering is not necessary:  The
frequencies are actually complex with imaginary parts that produce
damping on timescales $\alt 1$
second (when a servo is introduced to control
an instability). Therefore, the influence of $\hat x_o$ and $\hat p_o$
on the output flux operator $\hat{\cal N}(t)$ damps out quickly, and
correspondingly (see the end of Sec.\ \ref{sec:PhotonFluxCommutator}), 
the influence of the test-mass initial state
on the output data train damps out quickly without any filtering.

\section{Conclusions}
\label{sec:Conclusions}

To reiterate:  In an interferometer (and many other force-measuring devices),
the output signal is encoded in the photon number flux operator $\hat {\cal
N}(t)$ of a light beam, which is converted into discrete photon number samples
$\hat N_j$ by a photodetector and electronics.  These outputs have vanishing
commutators $[\hat {\cal N}(t), \hat {\cal N}(t')] = 0$ and $[\hat N_j,\hat
N_k] = 0$ and thus can be thought of as classical quantities.  These outputs
are linear in the initial test-mass position $\hat x_o$ and momentum $\hat p_o$
and involve no other test-mass variables.  The output commutators manage to
vanish because the photon back-action noise and photon shot noise have
commutators that cancel those of $\hat x_o$ and $\hat p_o$.  

In the output
$\hat {\cal N}(t)$ of 
any interferometer with pendular dynamics, 
$\hat x_o$ and $\hat p_o$ appear only at the pendular frequency $\omega_m/2\pi
\sim 1$~Hz, and all influences of $\hat x_o$ and $\hat p_o$ (including
all influences of the test-mass initial state) are removed completely
from the data by the high-pass filtering that is
routine for interferometers.  For other types of interferometers,
with different test-mass dynamics, other data filtering procedures will
remove the influence of $\hat x_o$ and $\hat p_o$ and the test-mass initial
state --- and in some cases (e.g., a signal-recycled interferometer)
no filtering is needed at all.

This complete removal of all influence of $\hat x_o$ and $\hat p_o$ from the
filtered data implies the answers to the three questions posed in the
introduction of this paper (Sec.\ \ref{sec:Questions}): (i) The test-mass
quantum mechanics has no influence on the interferometer's noise; the only
quantum noise is that arising from the light.  (ii) Therefore, when analyzing
a candidate interferometer design, one need not worry about the test-mass
quantum mechanics, except for using it to feed the gravity-wave signal and the
back-action noise through the test mass to the photon-flux output.  (iii)
Similarly, when conceiving new designs for interferometers, one need not worry
about the test-mass quantum mechanics 
--- except for devising appropriate
data filters to remove $\hat x_o$ and $\hat p_o$ from the data. 

\section*{Acknowledgments}
For helpful advice or email correspondence, 
we thank  Orly Alter,
Alessandra Buonanno,
Carlton Caves, Yanbei Chen, Crispin Gardiner, William
Unruh, 
Yoshihisa Yamamoto, 
and the members of the 
1998--99 Caltech QND Reading Group, most especially
Constantin Brif, Bill Kells, Jeff Kimble, Yuri Levin and John Preskill.  This
research was supported in part by NSF grants PHY--9503642, 
PHY--9900776, PHY-0098715, and PHY--0099568, 
by the Russian Foundation for Fundamental Research grants
\#96-02-16319a and \#97-02-0421g, and (for VBB, FYaK and SPV) by the
NSF through Caltech's Institute for Quantum Information.

\appendix

\section{Triple Measurement in the Schroedinger Picture}
\label{app:TripleMeasurement}

In this appendix we present a Schroedinger-picture analysis of the
most important of this paper's pedagogical thought experiments
(Sec.\ \ref{sec:VonNeumann}):  a
triple measurement of the position of a free test mass, using three 
independent meters, with the goal of determining the mean classical 
force $\bar F$ acting on the test mass without any contaminating
noise whatsoever from the test mass's initial state.  Our analysis
will proceed in three steps: (i) an analysis of one of the 
position measurements (any one of the three), 
Sec.\ \ref{sec:single_measurement}; (ii) [relying on step (i)]
a derivation of the probability
density $W(\tilde Q_0, \tilde Q_1, \tilde Q_2)$ for the outcome of
the triple measurement procedure, Sec.\ \ref{sec:triple_measurement};
and (iii) a use of this probability density to show that the 
combination $\tilde R \equiv \tilde Q_0 - 2 \tilde Q_1 + \tilde Q_2$
of the measurement results contains the desired information about
$\bar F$ uncontaminated by any noise from the test-mass initial state,
Sec.\ \ref{sec:statistics}.

\subsection{Single position measurement}
\label{sec:single_measurement}

Let $|{\Psi}\rangle$ be the state of the test mass before the
measurement and
\begin{equation}
  |{\psi}\rangle = 
\displaystyle\int_{-\infty}^{\infty}
\psi(Q)|{Q}\rangle\,dQ
\label{psiQ}
\end{equation}
be the initial state of the meter,
where the meter's eigenstates are normalized by
\begin{equation}
\langle Q' | Q \rangle = \delta(Q-Q')\;.
\label{Qnormalize}
\end{equation}
We leave the test-mass state $|{\Psi}\rangle$
completely unspecified since our goal is to show that it has no
influence at all on the measurement outcome. For concreteness
we specify 
the meter's initial wave function
$\psi(Q)$ to be Gaussian:
\begin{equation}\label{psi}
  \psi(Q) = \frac{1}{\sqrt{\sqrt{2\pi}\,\Delta_Q}}\,\exp\left[
    -\frac{Q^2}{2\Delta_Q^2}\left(\frac12-\frac{i\Delta_{QP}}{\hbar}\right)
  \right] \;.
\end{equation}
Here $\Delta_Q$ (denoted $\Delta Q^{\rm before}$ in the text) 
is the initial variance of $Q$ and
\begin{equation}
  \Delta_{QP} = \frac{\langle{\hat Q\hat P + \hat P\hat Q}\rangle}2
\end{equation}
is the initial cross correlation of the meter's position and momentum. For
this Gaussian initial state, the variance $\Delta_P$ of the meter's momentum 
(denoted $\Delta P^{\rm before}$ in the text) is
given by the 
minimum-uncertainty 
relation
\begin{equation}
  \Delta_Q^2\Delta_P^2 - \Delta_{PQ}^2 = \frac{\hbar^2}{4} \;.
\end{equation}

The first stage of the measurement process is the
interaction of the test mass and the meter.  In the Schroedinger Picture
this interaction puts the meter and test mass into the entangled state
\begin{equation}
  \hat U|{\psi}\rangle |{\Psi}\rangle \;,
\end{equation}
where
\begin{equation}\label{U}
  \hat U = \exp{\left(\frac{i\hat x\hat P}{\hbar}\right)}
\end{equation}
is the evolution operator associated with the interaction (delta
function) part of the Hamiltonian (\ref{Hamiltonian}). 

The next stage is a precise measurement of the meter's generalized 
position $\hat Q$.   This measurement disentangles the quantum states
of the test mass and meter:  the meter gets reduced to
the eigenstate $|{\tilde Q }\rangle$ of $\hat Q$, where $\tilde Q$ is the
$c$-number obtained as result of this measurement, and the test mass gets
reduced to the state
\begin{equation}\label{redstate}
  \frac{\langle{\tilde Q}|\hat U|{\psi}\rangle |{\Psi}\rangle}{\sqrt{W(\tilde Q)}}
  = \frac{\hat\Omega(\tilde Q)|{\Psi}\rangle}{\sqrt{W(\tilde Q)}} \;,
\end{equation}
where
\begin{equation}\label{Omega}
  \hat\Omega(\tilde Q) = \langle{\tilde Q}|\hat U|{\psi}\rangle
\end{equation}
is the reduction operator describing the entire two-stage measurement
procedure, and 
\begin{equation}\label{W}
  W(\tilde Q) = \langle{\Psi}|\hat\Omega^\dagger(\tilde Q)
    \hat\Omega(\tilde Q)|{\Psi}\rangle
\end{equation}
is the probability density for obtaining the result $\tilde Q$.

An explicit form for the reduction operator can be obtained by
substituting 
Eqs.\  
(\ref{psiQ}),
(\ref{psi}) and (\ref{U}) into Eq.\
(\ref{Omega}); the result is:
\begin{eqnarray}
  \Omega(\tilde Q) &=& \langle{\tilde Q}|
    \exp{\left(\frac{i\hat x\hat P}{\hbar}\right)}
\displaystyle\int_{-\infty}^{\infty}
\psi(Q)|{Q}\rangle\,dQ \nonumber\\
  &=& \langle{\tilde Q}|
\displaystyle\int_{-\infty}^{\infty}
|{x}\rangle\langle{x}|\,\psi(Q)|{Q-x}\rangle\,dx\,dQ \nonumber\\
  &=& 
\displaystyle\int_{-\infty}^{\infty}
|{x}\rangle\langle{x}|\psi(\tilde Q+x)\,dx  \nonumber\\
  &=& \frac{1}{\sqrt{\sqrt{2\pi}\,\Delta_Q}}\,\exp\left[
      -\frac{(\tilde Q + \hat x)^2}{2\Delta_Q^2}
      \left(\frac12-\frac{i\Delta_{QP}}{\hbar}\right)
  \right] \;,
\nonumber\\
\label{OmegaNorm}
\end{eqnarray}
where we have used the shift-operator relation 
$e^{i\hat x \hat P / \hbar} |Q\rangle = |Q-\hat x\rangle 
= \int_{-\infty}^{\infty} dx |x\rangle\langle x|Q-x\rangle$
and the relation $\langle \tilde Q | Q-x\rangle = 
\delta(Q-x-\tilde Q)$.

We will need below the following formulae (some are evident, and for
the others we provide outlines of the proofs):

\FL
\begin{equation}\label{int_none}
\displaystyle\int_{-\infty}^{\infty}
\hat\Omega^\dagger(\tilde Q)\hat\Omega(\tilde Q)\,d\tilde Q = 1 \;,
\end{equation}

\FL
\begin{equation}\label{int_Q}
\displaystyle\int_{-\infty}^{\infty}
\hat\Omega^\dagger(\tilde Q)\hat\Omega(\tilde Q)\,
    \tilde Q\,d\tilde Q = - \hat x \;,
\end{equation}

\FL
\begin{equation}\label{int_Q2}
\displaystyle\int_{-\infty}^{\infty}
\hat\Omega^\dagger(\tilde Q)\hat\Omega(\tilde Q)\,
    \tilde Q^2\,d\tilde Q = \hat x^2 + \Delta_Q^2 \;,
\end{equation}

\FL
\begin{equation}\label{int_x}
\displaystyle\int_{-\infty}^{\infty}
\hat\Omega^\dagger(\tilde Q)\hat x^n\hat\Omega(\tilde Q)\,
    d\tilde Q = \hat x^n \qquad (n=0,1,\dots) \;,
\end{equation}

\FL
\begin{equation}\label{int_x0}
\displaystyle\int_{-\infty}^{\infty}
\hat\Omega^\dagger(\tilde Q)\hat x\hat\Omega(\tilde Q)\,
    \tilde Q\,d\tilde Q = -\hat x^2 \;,
\end{equation}

\FL
\begin{eqnarray}\label{int_p}
\lefteqn{
\displaystyle\int_{-\infty}^{\infty}
\hat\Omega^\dagger(\tilde Q)\hat p\hat\Omega(\tilde Q)\,d\tilde Q}\quad 
\nonumber\\
&&= 
\displaystyle\int_{-\infty}^{\infty}
\hat\Omega^\dagger(\tilde Q)\left(
      \hat\Omega(\tilde Q)\hat p + \left[\hat p,\hat\Omega(\tilde Q)\right]
    \right)\,d\tilde Q \nonumber\\
&&= 
\displaystyle\int_{-\infty}^{\infty}
\hat\Omega^\dagger(\tilde Q)\hat\Omega(\tilde Q)\,
      d\tilde Q\,\hat p
    - i\hbar
\displaystyle\int_{-\infty}^{\infty}
\hat\Omega(\tilde Q)
        \frac{d\hat\Omega^\dagger(\tilde Q)}{d\hat x}\,d\tilde Q\nonumber\\
&&= \hat p \;,
\end{eqnarray}

\FL
\begin{eqnarray}
\lefteqn{\displaystyle\int_{-\infty}^{\infty}
\hat\Omega^\dagger(\tilde Q)\hat p^2\hat\Omega(\tilde Q)\,d\tilde Q}
\nonumber\\
  &=& 
\displaystyle\int_{-\infty}^{\infty}
\left(
      \hat p\hat\Omega^\dagger(\tilde Q)
      + \left[\hat\Omega^\dagger(\tilde Q),\hat p\right]
    \right)\left(
      \hat\Omega(\tilde Q)\hat p + \left[\hat p,\hat\Omega(\tilde Q)\right]
    \right)\,d\tilde Q \nonumber\\
  &=& \hat p
\displaystyle\int_{-\infty}^{\infty}
\hat\Omega^\dagger(\tilde Q)\hat\Omega(\tilde Q)\,
      d\tilde Q\,\hat p
    + \hbar^2
\displaystyle\int_{-\infty}^{\infty}
\frac{d\hat\Omega^\dagger(\tilde Q)}{d\hat x}
        \frac{d\hat\Omega(\tilde Q)}{d\hat x}\,d\tilde Q  \nonumber\\
  &=& \hat p^2 + \frac1{\Delta_Q^2}\left(\frac{\hbar^2}4+\Delta_{QP}^2\right)
  = \hat p^2 + \Delta_P^2 \;,
\label{int_1}
\end{eqnarray}

\FL
\begin{eqnarray}
\lefteqn{\displaystyle\int_{-\infty}^{\infty}
\hat\Omega^\dagger(\tilde Q)\hat p\hat\Omega(\tilde Q)\,
    \tilde Q\,d\tilde Q} \nonumber\\
  &=& 
\displaystyle\int_{-\infty}^{\infty}
\hat\Omega^\dagger(\tilde Q)\left(
      \hat\Omega(\tilde Q)\hat p + \left[\hat p,\hat\Omega(\tilde Q)\right]
    \right)\,\tilde Q\,d\tilde Q \nonumber\\
  &=& 
\displaystyle\int_{-\infty}^{\infty}
\hat\Omega^\dagger(\tilde Q)\hat\Omega(\tilde Q)\,
      \tilde Q\,d\tilde Q\,\hat p
    - i\hbar
\displaystyle\int_{-\infty}^{\infty}
\hat\Omega(\tilde Q)
        \frac{d\hat\Omega^\dagger(\tilde Q)}{d\hat x}\,\tilde Q\,d\tilde Q 
\nonumber\\
  &=& -\hat x\hat p + i\hbar\left(\frac12 - \frac{\Delta_{QP}}{\hbar}\right)
  = - \frac{\hat x\hat p + \hat p\hat x}2 + \Delta_{QP} \;,
\end{eqnarray}

\FL
\begin{equation}\label{int_xp}
\displaystyle\int_{-\infty}^{\infty}
\hat\Omega^\dagger(\tilde Q)(\hat x\hat p+\hat p\hat x)
    \hat\Omega(\tilde Q)\,d\tilde Q
  = \hat x\hat p+\hat p\hat x \;.
\label{int_pQ}
\end{equation}

\subsection{The triple measurement procedure}
\label{sec:triple_measurement}

The triple measurement procedure described in Sec.\ \ref{sec:VNThought} of the
text consists of the following five stages.

\begin{enumerate}

\item

An initial position measurement of the type we have just analyzed, 
using meter number 0. This
measurement reduces the test mass's wave function to 
\begin{equation}
  \frac{\hat\Omega_0(\tilde Q_0)|{\Psi}\rangle}{\sqrt{W_0(\tilde Q_0)}}\;
\end{equation}
[Eq.\ (\ref{redstate})],
where $\hat\Omega_0(\tilde Q_0)$ is the reduction operator 
[Eq.\ (\ref{Omega})],
and $\tilde Q_0$ is
the result of this measurement. The probability density for obtaining this 
result is equal to
\begin{equation}
  W_0(\tilde Q_0) = \langle{\Psi}|
\hat\Omega_0^\dagger(\tilde Q_0) \hat\Omega_0(\tilde Q_0)
|{\Psi}\rangle \;
\end{equation}
[Eq.\ (\ref{W})].

\item

Free evolution of the test mass during the time $\tau$. Denoting the 
corresponding
evolution operator by $\hat{\cal U}_0$, the test-mass wave function 
after this stage is given by
\begin{equation}
  \frac{\hat{\cal U}_0\hat\Omega_0(\tilde Q_0)
    |{\Psi}\rangle}{\sqrt{W_0(\tilde Q_0)}}\;.
\end{equation}

\item

Second position measurement of the same type as in the first stage,
but using a new meter, number 1. The measurement result is denoted $\tilde
Q_1$, the
reduction operator is $\Omega_1(\tilde Q_1)$, and the measurement reduces
the test-mass state to 
\begin{equation}
  \frac{\hat\Omega_1(\tilde Q_1)\hat{\cal U}_0\hat\Omega_0(\tilde Q_0)
    |{\Psi}\rangle}{\sqrt{W_1(\tilde Q_0,\tilde Q_1)}}\;,
\end{equation}
where
\begin{eqnarray}
&&  W_1(\tilde Q_0,\tilde Q_1) \nonumber\\
&&= \langle {\Psi}|
    \hat\Omega_0^\dagger(\tilde Q_0)\hat{\cal U}_0^\dagger
    \hat\Omega_1^\dagger(\tilde Q_1)\hat\Omega_1(\tilde Q_1)
    \hat{\cal U}_0\hat\Omega_0(\tilde Q_0)
  |{\Psi}\rangle
\end{eqnarray}
is the joint probability disrtibution for the first two measurement
results, $\tilde Q_0$ and $\tilde Q_1$.

\item

Second free evolution of the test mass with the evolution operator
$\hat{\cal U}_1$. After this stage the test-mass wave function is
\begin{equation}
  \frac{\hat{\cal U}_1\hat\Omega_1(\tilde Q_1)
    \hat{\cal U}_0\hat\Omega_0(\tilde Q_0)
    |{\Psi}\rangle}{\sqrt{W_1(\tilde Q_0,\tilde Q_1)}}\;.
\end{equation}

\item

Finally, a third position measurement using a new meter, number 2,  
with the result $\tilde Q_2$. After this measurement the test-mass state is
\begin{equation}
  \frac{\hat\Omega_2(\tilde Q_2) \hat{\cal U}_1\hat\Omega_1(\tilde Q_1)
    \hat{\cal U}_0\hat\Omega_0(\tilde Q_0)
    |{\Psi}\rangle}{\sqrt{W_2(\tilde Q_0,\tilde Q_1,\tilde Q_2)}}\;,
\end{equation}
where
\FL
\begin{eqnarray}
\lefteqn{  W_2(\tilde Q_0,\tilde Q_1,\tilde Q_2) =
  \langle{\Psi}|
    \hat\Omega_0^\dagger(\tilde Q_0)\hat{\cal U}_0^\dagger
    \hat\Omega_1^\dagger(\tilde Q_1)\hat{\cal U}_1^\dagger
    \hat\Omega_2^\dagger(\tilde Q_2) 
} \quad\quad\quad
\nonumber\\ 
&&\times
    \hat\Omega_2(\tilde Q_2)
    \hat{\cal U}_1\hat\Omega_1(\tilde Q_1)
    \hat{\cal U}_0\hat\Omega_0(\tilde Q_0)
  |{\Psi}\rangle \;.
\label{W_2}
\end{eqnarray}
is the joint probability 
distribution for all three measurement outcomes.

\end{enumerate}

Equation (\ref{W_2}) is the principal result of this subsection. We shall
use it to study the statistics of the measurement outcomes.  In that study
we shall need the following expression for each of the three reduction
operators
[Eq.\ (\ref{OmegaNorm})]:
\FL
\begin{eqnarray}
\lefteqn{
  \hat\Omega_s(\tilde Q_s)
} \nonumber \\
&& = \frac{1}{\sqrt{\sqrt{2\pi}\,\Delta_{Q\,s}}}\,
    \exp\left[
      -\frac{(\tilde Q_s + \hat x)^2}{2\Delta_{Q\,s}^2}
      \left(\frac12-\frac{i\Delta_{QP\,s}}{\hbar}\right)
  \right] ,
\end{eqnarray}
where $s=1,2,3$.

\subsection{Statistics of the measurement results}
\label{sec:statistics}

If an explicit form for the initial wave function $|{\Psi}\rangle$
were specified, then the probability density (\ref{W_2}) 
could be calculated directly. However,
that calculation would be very cumbersome, the final result
would be quite complicated, and we have no need for it.  Our final
goal is not to study $W_2$, but rather to 
analyze the statistics of the quantity $\tilde R =
\tilde Q_0 - 2 \tilde Q_1 + \tilde Q_2$, which the experimenter computes
from the three measurement outcomes $\tilde Q_s$ after the triple
measurement procedure is complete.  Specifically, we wish to verify
the results of the text's Heisenberg-picture analysis:  (i) That the
mean value of $\tilde R$ over a large number of experiments is 
$\langle{\tilde R}\rangle = 
(-\tau^2/\mu)\bar F$, where $\tau$ is the time between each pair of
measurements, $\mu$ is the mass of the test mass, and $\bar F$ 
is the mean force that acts on the test mass [Eqs.\ (\ref{xiVal})  
and (\ref{hatR}) of the text].
(ii) That the variance of $\tilde R$ (and thence of the measured value of
$\bar F$) {\it is independent of the
test-mass initial state} $|{\Psi}\rangle$, and is given by 
Eq.\ (\ref{DeltabarF})
when the meters' individual initial states have no position-momentum
correlations, $\Delta_{QP\,s}=0$, and can be made to vanish by a
clever, ``squeezed'' choice of the meters' initial states. 
\paragraph{Mean value.}

The mean value of $\tilde R$ over a large number of experiments is 
determined by the joint probability distribution $W_3$ for the measurement
outcomes:
\FL
\begin{eqnarray}
\langle{\tilde R}\rangle 
&=& \langle{\tilde Q_0-2\tilde Q_1+\tilde Q_2}\rangle
\nonumber\\
  &=& 
\displaystyle\int_{-\infty}^{\infty}
(\tilde Q_0-2\tilde Q_1+\tilde Q_2)
      W_2(\tilde Q_0,\tilde Q_1,\tilde Q_2)\,
      d\tilde Q_0d\tilde Q_1d\tilde Q_2 . \nonumber\\
\label{mean_0}
\end{eqnarray}
Using Eqs.\ (\ref{int_none}), (\ref{int_Q}), we bring this into the form
\FL
\begin{eqnarray}
\lefteqn{
  \langle{\tilde R}\rangle 
  = 
\displaystyle\int_{-\infty}^{\infty}
\langle{\Psi}|
      \hat\Omega_0^\dagger(\tilde Q_0)\hat{\cal U}_0^\dagger
      \hat\Omega_1^\dagger(\tilde Q_1)\hat{\cal U}_1^\dagger
} \nonumber\\
&&\quad \times
      \left(\tilde Q_0-2\tilde Q_1-\hat x\right)
      \hat{\cal U}_1\hat\Omega_1(\tilde Q_1)
      \hat{\cal U}_0\hat\Omega_0(\tilde Q_0)
    |{\Psi}\rangle\,d\tilde Q_0d\tilde Q_1 \;.
\nonumber\\
\label{mean_1}
\end{eqnarray}
Taking into account that
\begin{eqnarray}
  {\cal U}_1^\dagger{\cal U}_1 &=& 1 \;, \label{U_1_none} \\
  {\cal U}_1^\dagger\hat x{\cal U}_1
    &=& x + \frac{\hat p\tau}\mu + x_{F\,1}\;, \label{U_1_x}
\end{eqnarray}
where $\mu$ is the mass of the test mass and
\begin{equation}
  x_{F\,1} = \frac1\mu\,\int_\tau^{2\tau}(2\tau-t)F(t)\,dt
\end{equation}
is the displacement of the test mass during stage 4 (the second interval of
free evolution) caused by the external force $F(t)$, expression (\ref{mean_1})
can be further reduced to the form
\FL
\begin{eqnarray}
  \langle{\tilde R}\rangle 
  &=& 
\displaystyle\int_{-\infty}^{\infty}
\langle{\Psi}|
      \hat\Omega_0^\dagger(\tilde Q_0){\cal U}_0^\dagger
      \hat\Omega_1^\dagger(\tilde Q_1) 
\left(
        \tilde Q_0-2\tilde Q_1
\right.
\nonumber\\
&&\left. 
\quad -\hat x - \frac{\hat p\tau}\mu - x_{F\,1}
      \right) 
      \hat\Omega_1(\tilde Q_1)\hat{\cal U}_0\hat\Omega_0(\tilde Q_0)
    |{\Psi}\rangle\,d\tilde Q_0d\tilde Q_1 \;.
\nonumber\\
\label{mean_2}
\end{eqnarray}
The next calculations are just a repetition of the previous ones, with 
only the addition of Eqs.\ (\ref{int_x}), (\ref{int_p}) and 
\begin{eqnarray}
  {\cal U}_0^\dagger\hat x{\cal U}_0
    = x + \frac{\hat p\tau}\mu + x_{F\,0} \;, \label{U_0_x} \\
  {\cal U}_0^\dagger\hat p{\cal U}_0 = p + p_{F\,0}\;, \label{U_0_p}
\end{eqnarray}
where
\begin{eqnarray}
  x_{F\,0} = \frac1\mu\,\int_0^\tau(\tau-t)F(t)\,dt  \;, \\
  p_{F\,0} = \int_0^\tau F(t)\,dt  \;.
\end{eqnarray}
They give:
\FL
\begin{eqnarray}
\lefteqn{
  \langle{\tilde R}\rangle
  = 
\displaystyle\int_{-\infty}^{\infty}
\langle{\Psi}|
      \hat\Omega_0^\dagger(\tilde Q_0){\cal U}_0^\dagger
      \left(
        \tilde Q_0 + 2\hat x - \hat x - \frac{\hat p\tau}\mu - x_{F\,1}
      \right)
}
\nonumber\\
&&\quad\times
      {\cal U}_0\hat\Omega_0(\tilde Q_0)
    |{\Psi}\rangle\,d\tilde Q_0 \nonumber\\
  &=& 
\displaystyle\int_{-\infty}^{\infty}
\langle{\Psi}|
      \hat\Omega_0^\dagger(\tilde Q_0){\cal U}_0^\dagger
\nonumber\\
&&\quad\times
      \left(
        \tilde Q_0 + \hat x + x_{F\,0} - \frac{p_{F\,0}\tau}\mu - x_{F\,1}
      \right)
      \hat\Omega_0(\tilde Q_0)
    |{\Psi}\rangle\,d\tilde Q_0 \nonumber\\
  &=& \langle{\Psi}|\left(
      x_{F\,0} - \frac{p_{F\,0}\tau}\mu - x_{F\,1}
    \right)|{\Psi}\rangle
  = x_{F\,0} - \frac{p_{F\,0}\tau}\mu - x_{F\,1} \nonumber\\
  &=& -\frac1\mu\int_0^{2\tau}(\tau-|t-\tau|)F(t) \,dt  
  \equiv -\frac{\tau^2}\mu \bar F\;.
\label{mean_4}
\end{eqnarray}
This agrees with the Heisenberg-picture prediction
[Eqs.\ (\ref{xiVal}) and (\ref{hatR}) of the text, where we must note
that the meters' initial states have $\langle{Q_s}\rangle = \langle{P_s}\rangle=0$].

\paragraph{Variance}

The mean square value of the measurement outcome $\tilde R$ over a 
large number of experiments is given by
\FL
\begin{eqnarray}
\lefteqn{
 \langle{\tilde R^2}\rangle 
=  \langle{(\tilde Q_0-2\tilde Q_1+\tilde Q_2)^2}\rangle
}
\nonumber\\
&& \;\;\;= 
\displaystyle\int_{-\infty}^{\infty}
(\tilde Q_0-2\tilde Q_1+\tilde Q_2)^2
      W_2(\tilde Q_0,\tilde Q_1,\tilde Q_2)\,
      d\tilde Q_0d\tilde Q_1d\tilde Q_2 \;.
\nonumber\\
\label{sqear_0}
\end{eqnarray}
Using Eqs.\ (\ref{int_none})--(\ref{int_xp}), (\ref{U_1_none}),
(\ref{U_1_x}), (\ref{U_0_x}), and (\ref{U_0_p}), we obtain:
\FL
\begin{eqnarray}\label{sqear_1}
\lefteqn{
 \langle{\tilde R^2}\rangle = \langle{(\tilde Q_0-2\tilde Q_1+\tilde Q_2)^2}\rangle
}
 \nonumber\\
  &&= 
\displaystyle\int_{-\infty}^{\infty}
\langle{\Psi}|
      \hat\Omega_0^\dagger(\tilde Q_0)\hat{\cal U}_0^\dagger
      \hat\Omega_1^\dagger(\tilde Q_1)\hat{\cal U}_1^\dagger
      \left[
        (\tilde Q_0-2\tilde Q_1-\hat x)^2 + \Delta_Q^2
      \right] \nonumber\\
  &&\times    \hat{\cal U}_1\hat\Omega_1(\tilde Q_1)
      \hat{\cal U}_0\hat\Omega_0(\tilde Q_0)
    |{\Psi}\rangle\,d\tilde Q_0d\tilde Q_1 \nonumber\\
  &&=
\displaystyle\int_{-\infty}^{\infty}
\langle{\Psi}|
      \hat\Omega_0^\dagger(\tilde Q_0){\cal U}_0^\dagger
      \hat\Omega_1^\dagger(\tilde Q_1)
\nonumber\\
&&\times
      \left[
        \left(
          \tilde Q_0 - 2\tilde Q_1 - \hat x - \frac{\hat p\tau}m - x_{F\,1}
        \right)^2 + \Delta_{Q\,2}^2
      \right] \nonumber\\
  &&\times
      \hat\Omega_1(\tilde Q_1)\hat{\cal U}_0\hat\Omega_0(\tilde Q_0)
    |{\Psi}\rangle\,d\tilde Q_0d\tilde Q_1 \nonumber\\
  &&=
\displaystyle\int_{-\infty}^{\infty}
\langle{\Psi}|
      \hat\Omega_0^\dagger(\tilde Q_0){\cal U}_0^\dagger
      \left[
        \left(\tilde Q_0 + \hat x - \frac{\hat p\tau}m - x_{F\,1}\right)^2
      + 4\Delta_{Q\,1}^2 
\right.
\nonumber\\
&&
\left.
         + \frac{4\Delta_{QP\,1}\tau}m + 
\left({\Delta_{P\,1} \tau\over\mu}\right)^2
         + \Delta_{Q\,2}^2
      \right] 
      \hat{\cal U}_0\hat\Omega_0(\tilde Q_0)
    |{\Psi}\rangle\,d\tilde Q_0 \nonumber\\
  &&=
\displaystyle\int_{-\infty}^{\infty}
\langle{\Psi}|
      \hat\Omega_0^\dagger(\tilde Q_0)
      \left[
        \left(
          \tilde Q_0 + \hat x + x_{F\,0} - \frac{p_{F\,0}\tau}m - x_{F\,1}
        \right)^2
\right.
\nonumber\\
&&
\left.
          + 4\Delta_{Q\,1}^2 + \frac{4\Delta_{QP\,1}\tau}m + 
\left({\Delta_{P\,1} \tau\over\mu}\right)^2
          + \Delta_{Q\,2}^2
      \right] 
      \hat\Omega_0(\tilde Q_0)
    |{\Psi}\rangle\,d\tilde Q_0 \nonumber\\
  &&= \langle{\Psi}|
      \left[
        \left(x_{F\,0} - \frac{p_{F\,0}\tau}m - x_{F\,1}\right)^2
        + \Delta_{Q\,0}^2
\right.
\nonumber\\
&&
\left.
        + 4\Delta_{Q\,1}^2 + \frac{4\Delta_{QP\,1}\tau}m + 
\left({\Delta_{P\,1} \tau\over\mu}\right)^2
        + \Delta_{Q\,2}^2
      \right]
    |{\Psi}\rangle \nonumber\\
  &&=
\langle{\tilde Q_0-2\tilde Q_1+\tilde Q_2}\rangle^2
    + \Delta_{Q\,0}^2
\nonumber\\
&&
    + 4\Delta_{Q\,1}^2 + \frac{4\Delta_{QP\,1}\tau}m + 
\left({\Delta_{P\,1} \tau\over\mu}\right)^2
    + \Delta_{Q\,2}^2
\end{eqnarray}
Subtracting off the square of the mean, $\langle{\tilde R}\rangle^2 = 
\langle{\tilde Q_0-2\tilde Q_1+\tilde Q_2}\rangle^2$, we obtain for the variance of
the computed quantity $\tilde R$, over many experiments,
\FL
\begin{eqnarray}
{\tau^4\over\mu^2}(\Delta \bar F)^2
&=& (\Delta\tilde R)^2 = \langle \hat R^2\rangle - \langle \hat
R\rangle^2
\nonumber\\
  &=& \Delta_{Q\,0}^2
    + 4\Delta_{Q\,1}^2 + \frac{4\Delta_{QP\,1}\tau}m + 
\left({\Delta_{P\,1} \tau\over\mu}\right)^2
    + \Delta_{Q\,2}^2 \;;
\nonumber\\
\label{DeltaR}
\end{eqnarray}
see Eq.\ (\ref{mean_4}) for the first equality.
{\it This variance is independent of the test-mass initial state} 
$|{\Psi}\rangle$,
in accord with prediction of the Heisenberg-picture analysis 
[passage following Eq.\ (\ref{GExpectation}) of the text].  When the three
meters are all prepared in ``naive'' initial
states, i.e.\ in states with uncorrelated generalized position $\hat Q_s$
and momentum $\hat P_s$, i.e.\ when $\Delta_{QP\,s} = 0$, then the
variance (\ref{DeltaR}) has the form that we deduced using the 
Heisenberg picture [Eq.\ (\ref{DeltabarF}) ].  When the meters are prepared in
the more clever ``squeezed'' manner, i.e.\ in near eigenstates of
$\hat Q_0$, $\hat Q_1^{\rm squeeze} = \hat Q_1 - \hat P_1\tau/2\mu$
and $\hat Q_2$, then the variance (\ref{DeltaR}) vanishes, 
in accord with the Heisenberg-picture prediction [passage following
Eq.\ (\ref{IdealSqueezed})].

\section{Linear measurements}
\label{app:LinearMeasurements}

An important feature of our pedagogical examples (Sec.\ \ref{sec:Pedagogy}),
and of measurements performed by interferometric gravitational-wave detectors,
is that they all are
{\it linear measurements} in the sense of Ref.\ \cite{QuantumMeasurement}; i.e.,
they all satisfy the following two conditions:

(i) {\it Linearity of the output:} The meter's output can be written 
as the sum of the operator for the test object's measured variable 
and the operator for the meter's 
additive noise [cf.\ Eq.\ (\ref{SimpleEqsA})],  
and the additive noise
does not depend on the initial state of the test object. 
Formally this sum is an operator, but it can be treated as 
a classical variable 
because it turns out to commute with itself at different times.

(ii) {\it Linearity of the back action:}  The measurement-induced 
perturbations of all the test-object observables that 
are involved in the measurement procedure can be described
by linear formulas similar to Eq.\ (\ref{SimpleEqsB}), and the
perturbations [e.g.\ the second term on the right side of (\ref{SimpleEqsB})]
do not depend on the initial state of the test object.

This second condition requires discussion:  The perturbations' independence
of the test-object initial state is particularly important when 
several test-object variables are measured consecutively --- for example,
if the same Heisenberg-Picture variable is measured quickly and repetitively 
at different moments of time as in our pedagogical examples 
(Sec.\ \ref{sec:Pedagogy}), or if a variable is measured continuously
as in a gravitational-wave detector (Sec.\ \ref{sec:IFOs}).
Suppose, for example, that the variable $\hat x_1$ is measured with precision
$\Delta x_1^{\rm meas}$ thereby perturbing, via back-action, some other variable 
$\hat x_2$.  Then the accuracy of a subsequent measurement of $\hat x_2$
will be 
constrained by the perturbation
\begin{equation}
  \Delta x_2^{\rm pert} = \frac{\hbar}{2\Delta x_1^{\rm meas}}
    |\langle [\hat x_1,\hat x_2]\rangle | \,. 
\end{equation}
Our condition (ii) of back-action linearity requires that this 
perturbation
not depend on the initial state of the test object.  A sufficient condition
for this is that the
commutator $[\hat x_1,\hat x_2]$ be
a $c$-number, and that this requirement be fulfilled for all the operators
involved in the measurement\footnote{
It can be shown that 
a slightly weaker condition is sufficient: second-order
commutation of all these operators,  $[\hat x_i,[\hat x_j,\hat x_k]] = 0$ for
all $i,j,k$.}

Linear measurements are closely related to linear systems (those for which
the equations of motion for the generalized coordinates and momenta are linear;
for example, a free mass and a harmonic oscillator) because the 
commutators of such systems' coordinates and momenta are $c$-numbers.

In {\it non}linear measurements 
(e.g.\ measurements of a particle in a double-welled
potential), some very strange phenomena can arise, for example the quantum
Zeno effect.

Strictly speaking, all real meters are nonlinear. However, in most cases they
can be regarded as linear to high accuracy. For example,  
if one measures displacements of a mirror of a Fabry-Perot cavity by 
monitoring the phase of light that passes through the cavity (as is
done in LIGO), then the
measurements are linear so long as the displacements are much smaller than
the width of a cavity resonance, i.e.\ much smaller than $\lambda/{\cal F}$
where $\lambda$ is the wavelength of the light and $\cal F$ is the cavity 
finesse.

If, by contrast, the displacements are comparable to or much larger than 
$\lambda/{\cal F}$, then the measurements are strongly nonlinear.  
An example is
a proposed {\it null-detector} technique \cite{NullDetector}
for measuring the phase of a mechanical oscillator, in which the
oscillating mass is an end mirror of a Fabry-Perot cavity, and the times
at which the mirror passes through cavity-resonant positions are measured with
high accuracy by the cavity's momentary transmissivity.  These measurements
are highly nonlinear because, in the proposed design, 
not only are the mirror displacements large compared to the cavity's
linearity regime, $\lambda/{\cal F}$; the mechanical oscillator's 
amplitude of zero-point 
oscillations $\delta x_{\rm zp}$ is also large compared to 
$\lambda/{\cal F}$. 
State reduction plays an important role in this null detector's measurements:
it drives the mechanical oscillator into a squeezed-phase state, thereby
facilitating a high-precision monitoring of the oscillator's phase 
\cite{NullDetector}.  It would be instructive to analyze the use of this
highly nonlinear meter to monitor a classical force that acts on the
oscillator's mass.  Does the oscillator's initial quantum state influence
the accuracy of the monitoring?

Three properties of an interferometric gravitational-wave detector
(interferometric position meter) allow one to consider
it as linear with sufficiently high precision to justify the
linear analysis given in this paper.  
{\it First},
its test-mass mirrors can be regarded as free
masses (or as harmonic oscillators if significant electromagnetic rigidity 
exists in the system).
{\it Second}, its linearity
range $\lambda/{\cal F}\sim 10^{-6}$cm 
is much greater than the
wave-induced  displacements of the test masses ($\alt 10^{-15}$ cm). 
Hence, the signal phase shift
of the output optical beam depends linearly on the displacement. 
{\it Third}, the measurement of the photon flux out the dark port is virtually
equivalent to the measurement of the phase of the output beam because 
(i) the signal
phase shift is much less than one radian and (ii) the mean value of the 
amplitude of the optical pumping field is much larger than the quantum 
uncertainties of its quadrature amplitudes.

For a detailed presentation of the theory of linear measurements see
Chaps.\ 5 and 6 of Ref.\ \cite{QuantumMeasurement}.  For a detailed
application of this theory to interferometric gravitational-wave detectors
see Ref.\ \cite{BC3}.

\section{Vanishing self commutator of the photon number flux}
\label{app:VanishingCommutator}

For any light beam (or other electromagnetic wave with confined cross section),
the number flux operator at some chosen transverse plane (e.g. the entry to a
photodetector) is
\begin{equation}
\hat{\cal N}(t) = \int_0^\infty {d\omega\over2\pi} \int_0^\infty 
{d\omega'\over2\pi} \;
\hat a_\omega^\dag \hat a_{\omega'} \; e^{i(\omega-\omega')t}\;.
\label{calNa}
\end{equation}
Here $\hat a_\omega^\dag$ is the creation operator and $\hat a_\omega$ the
annihilation operator for photons of frequency $\omega$, and their commutators
are
\begin{equation}
[\hat a_\omega,\hat a_{\omega'}] = [\hat a_\omega^\dag,\hat a_{\omega'}^\dag] =
0\;,\quad
[\hat a_\omega,\hat a_{\omega'}^\dag] = 2\pi \delta(\omega-\omega')\;.
\label{aCommutators}
\end{equation}
It is straightforward to verify from
Eqs.\ (\ref{calNa}) and (\ref{aCommutators}) that
\begin{equation}
[\hat {\cal N}(t),\hat {\cal N}(t')] = 0\;.
\label{calNCommute}
\end{equation}

Although this result is completely general,
it is instructive to derive the vanishing self commutator for the
specialized type of light beam that is used in interferometers and other
force-measuring devices: a beam consisting of a monochromatic carrier
with frequency $\omega_o$ plus sidebands embodied in $\hat a_\omega$ and $\hat
a_\omega^\dag$.  In this case to high accuracy we can linearize in the product
of the carrier field and the side-band fields, obtaining for the relevant
(side-band) photon flux
\begin{equation}
\hat{\cal N}_1 (t) = \sqrt{\cal N}_0 
\left[ \hat a(t) + \hat a^{\dag}(t)\right]\;.
\label{calN2}
\end{equation}
Here [in the notation of Eqs.\ (\ref{ElectricField})--(\ref{EQuadratures})]
${\cal N}_0={A_0}^2$ is the carrier's photon flux and $\hat a(t)$, 
$\hat a^{\dag}(t)$ are the time-domain side-band annihilation
and creation
operators with commutation relations [time-domain versions of 
(\ref{aCommutators})]  
\begin{eqnarray}
[\hat a(t), && \hat a(t')] = 0 \;, \quad
[\hat a^{\dag}(t),\hat a^{\dag}(t')] =0\;, \nonumber\\
&&[ \hat a(t), \hat a^{\dag} (t')] 
= \delta(t-t')\;.
\label{calaTimeCommutator}
\end{eqnarray}
It is straightforward, using these commutation relations, 
to verify that
\begin{equation}
[{\cal N}_1(t),{\cal N}_1(t')] = 0\;.
\label{CalNCommuteZero}
\end{equation}

It is interesting to note that, although the photon number flux self commutes,
the energy flux (energy passing a fixed transverse surface per unit time)
\begin{equation}
\hat {\cal E}(t) =
\hbar \int_0^\infty d\omega \int_0^\infty d\omega' \;
\sqrt{\omega\omega'}\; \hat a_\omega^\dag
\hat a_{\omega'} \; e^{i\omega(t-t')}
\label{CalE}
\end{equation}
does {\it not} self-commute,
\begin{equation}
[\hat {\cal E}(t), \hat {\cal E}(t')] \ne 0\;.
\label{calENotCommute}
\end{equation}
This can be thought of as due to the energy-time uncertainty relation for
photons.
On the other hand, when (as in gravitational-wave interferometers) the light
consists of a monochromatic carrier plus signals encoded in side bands
with frequency $\Omega = \omega-\omega_o \ll \omega_o$, then for all
practical purposes, $\hat {\cal E}(t)$ {\it does} self commute.

\end{document}